\documentclass[usenatbib,twocolumn]{mn2e}
\usepackage{graphics,epsfig}
\usepackage{graphicx,times}
\usepackage[usenames,dvipsnames]{color}
\usepackage{bm,url}

\newcommand{\EQ}{\begin{equation}}
\newcommand{\EN}{\end{equation}}
\newcommand{\EQA}{\begin{eqnarray}}
\newcommand{\ENA}{\end{eqnarray}}
\newcommand{\eq}[1]{(\ref{#1})}

\newcommand{\Eq}[1]{Eq.~(\ref{#1})}

\newcommand{\Sec}[1]{Sect.~\ref{#1}}

\newcommand{\Fig}[1]{Fig.~\ref{#1}}
\newcommand{\FFig}[1]{Fig.~\ref{#1}}
\newcommand{\Figs}[2]{Figs.~\ref{#1} and \ref{#2}}

\newcommand{\Tab}[1]{Table~\ref{#1}}
\newcommand{\bra}[1]{\langle #1\rangle}

\newcommand{\mean}[1]{\overline #1}
\newcommand{\meanrho}{\overline{\rho}}

{}
{}
{}

\newcommand{\meanSSSS}{\overline{\mbox{\boldmath ${\mathsf S}$}} {}}

\newcommand{\meanSSS}{\overline{\mathsf{S}}}
{}
{}
{}
{}
{}
{}
{}
{}
\newcommand{\meanAA}{\overline{\mbox{\boldmath $A$}}{}}{}
\newcommand{\meanBB}{\overline{\mbox{\boldmath $B$}}{}}{}
{}
{}
{}
{}
{}
{}
{}
{}
\newcommand{\meanJJ}{\overline{\mbox{\boldmath $J$}}{}}{}
{}
\newcommand{\meanUU}{\overline{\mbox{\boldmath $U$}}{}}{}

{}
{}
{}

\newcommand{\meanB}{\overline{B}}

\newcommand{\meanD}{\overline{D}}

\newcommand{\meanU}{\overline{U}}

\newcommand{\meanUrms}{\overline{U}_{\rm rms}}
\newcommand{\meanBrms}{\overline{B}_{\rm rms}}
\newcommand{\meanp}{\overline{p}}

{}

{}
{}

\newcommand{\eee}{\hat{\mbox{\boldmath $e$}} {}}

\newcommand{\kk}{\bm{k}}

\newcommand{\xx}{\bm{x}}

\newcommand{\BB}{\bm{B}}

\newcommand{\uu}{\mbox{\boldmath $u$} {}}
\newcommand{\UU}{\mbox{\boldmath $U$} {}}

\newcommand{\JJ}{\mbox{\boldmath $J$} {}}

\newcommand{\AAA}{\mbox{\boldmath $A$} {}}

\newcommand{\ff}{\mbox{\boldmath $f$} {}}

\newcommand{\grav}{\mbox{\boldmath $g$} {}}
\newcommand{\nab}{\mbox{\boldmath $\nabla$} {}}
\newcommand{\OO}{\bm{\Omega}}

\newcommand{\oo}{\mbox{\boldmath $\omega$} {}}

\newcommand{\RRRR}{\mbox{\boldmath ${\sf R}$} {}}
\newcommand{\SSSS}{\mbox{\boldmath ${\sf S}$} {}}

\newcommand{\ii}{{\rm i}}

\newcommand{\DD}{{\rm D} {}}

\newcommand{\dd}{{\rm d} {}}

\newcommand{\const}{{\rm const}  {}}

\def \fh {\tilde{{\bm f}}}

\def\la{\mathrel{\mathchoice {\vcenter{\offinterlineskip\halign{\hfil
$\displaystyle##$\hfil\cr<\cr\sim\cr}}}
{\vcenter{\offinterlineskip\halign{\hfil$\textstyle##$\hfil\cr<\cr\sim\cr}}}
{\vcenter{\offinterlineskip\halign{\hfil$\scriptstyle##$\hfil\cr<\cr\sim\cr}}}
{\vcenter{\offinterlineskip\halign{\hfil$\scriptscriptstyle##$\hfil\cr<\cr\sim\cr}}}}}

\def\Fr{\mbox{\rm Fr}}

\def\Co{\mbox{\rm Co}}

\def\Pm{\mbox{\rm Pr}_M}
\def\Pmturb{\mbox{\rm Pr}_M^{\rm turb}}
\def\Rm{\mbox{\rm Re}_M}

\def\Rey{\mbox{\rm Re}}

\def\Co{\mbox{\rm Co}}

\def\cs{c_{\rm s}}
\def\qpz{q_{\rm p0}}
\def\qp{q_{\rm p}}

\def\betap{\beta_{\rm p}}

\def\kf{k_{\rm f}}

\def\urms{u_{\rm rms}}

\def\nuT{\nu_{\rm T}}

\def\etaT{\eta_{\rm T}}

\def\Beq{B_{\rm eq}}
\def\Beqz{B_{\rm eq0}}
\def\tautd{\tau_{\rm td}}

\def\half{{\textstyle{1\over2}}}

\def\onethird{{\textstyle{1\over3}}}

\newcommand{\etal}{et al.}
\newcommand{\yapj}[3]{ #1, {ApJ,} {#2}, #3}

\newcommand{\yapjl}[3]{ #1, {ApJ,} {#2}, #3}

\newcommand{\yan}[3]{ #1, {Astron.\ Nachr.,} {#2}, #3}

\newcommand{\yana}[3]{ #1, {A\&A,} {#2}, #3}

\newcommand{\ypfb}[3]{ #1, {Phys.\ Fluids B,} {#2}, #3}

\newcommand{\ysov}[3]{ #1, {Sov.\ Astron.,} {#2}, #3}
\newcommand{\ysovl}[3]{ #1, {Sov.\ Astron.\ Lett.,} {#2}, #3}
\newcommand{\yjetp}[3]{ #1, {Sov.\ Phys.\ JETP,} {#2}, #3}

\newcommand{\ymn}[3]{ #1, {MNRAS,} {#2}, #3}

\newcommand{\ysci}[3]{ #1, {Science,} {#2}, #3}
\newcommand{\yscia}[3]{ #1, {Sci.\ Adv. } {#2}, #3}

\newcommand{\ypre}[3]{ #1, {Phys.\ Rev.\ E,} {#2}, #3}

\newcommand{\ynjp}[3]{ #1, {NJP,} {#2}, #3}

\graphicspath{{./fig/}{./png/}}

\title{Sharp magnetic structures from dynamos with density stratification}
\author[S. Jabbari et al.]{
Sarah Jabbari$^{1,2,3}$
\thanks{E-mail: sarah.jabbari@monash.edu},
Axel Brandenburg$^{2,3,4,5}$,
Nathan Kleeorin $^{6,2}$ and
Igor Rogachevskii $^{6,2}$
\\
$^1$School of Mathematical Sciences and Monash Centre for Astrophysics, 
Monash University, Clayton, VIC 3800, Australia\\
$^2$Nordita, KTH Royal Institute of Technology and Stockholm University,
    Roslagstullsbacken 23, SE-10691 Stockholm, Sweden\\
$^3$Department of Astronomy, AlbaNova University Center,
    Stockholm University, SE-10691 Stockholm, Sweden\\
$^4$JILA and Department of Astrophysical and Planetary Sciences,
    Box 440, University of Colorado, Boulder, CO 80303, USA\\
$^5$Laboratory for Atmospheric and Space Physics,
    3665 Discovery Drive, Boulder, CO 80303, USA\\
$^6$Department of Mechanical Engineering, Ben-Gurion University of the Negev,
    PO Box 653, Beer-Sheva 84105, Israel\\
}

\begin{document}
\date{\today,~ $ $Revision: 1.244 $ $, DOI: 10.1093/mnras/stx148}
\pagerange{2753--2765} \volume{467} \pubyear{2017}

\maketitle

\label{firstpage}

\begin{abstract}
Recent direct numerical simulations (DNS) of large-scale turbulent
dynamos in strongly stratified layers have resulted in surprisingly
sharp bipolar structures at the surface.
Here we present new DNS of helically and non-helically forced turbulence
with and without rotation and compare with corresponding mean-field
simulations (MFS) to show that these structures are a generic outcome
of a broader class of dynamos in density-stratified layers.
The MFS agree qualitatively with the DNS, but the period of oscillations
tends to be longer in the DNS.
In both DNS and MFS, the sharp structures are produced by
converging flows at the surface and might be driven
in nonlinear stage of evolution by the Lorentz force
associated with the large-scale dynamo-driven magnetic field
if the dynamo number is at least 2.5 times supercritical.
\end{abstract}
\begin{keywords}
dynamo -- turbulence -- sunspots
\end{keywords}

\section{Introduction}

Active regions appear at the solar surface as bipolar patches with
a sharply defined polarity inversion line in between.
Bipolar magnetic structures are generally associated with buoyant magnetic
flux tubes that are believed to pierce the surface \citep{Par55}.
Furthermore, \cite{Par75} proposed that only near the bottom of the
convection zone the large-scale field can evade magnetic buoyancy losses
over time scales comparable with the length of the solar cycle.
This led many authors to study the evolution of magnetic flux tubes
rising from deep within the convection zone to the surface
\citep{Caligari,Fan01,Fan08,JB09}.
Shortly before flux emergence, however,
the rising flux tube scenario would predict flow speeds
that exceed helioseismically observed limits \citep{Birch16}.
Moreover, the magnetic field expands and weakens significantly during
its buoyant ascent.
Therefore, some type of reamplification of magnetic field structures
near the surface appears to be necessary.

The negative effective magnetic pressure instability (NEMPI) may be one such
mechanism of field reamplification. It has been intensively studied
both analytically \citep{KRR89,KRR90,KMR93,KMR96,KR94,RK07} and numerically
using direct numerical simulations (DNS) and mean-field simulations (MFS)
\citep{BKR10,BKKMR11,BKKR12,BKR13,BGJKR14},
see also the recent review by \cite{BRK16}.
The reamplification mechanism of magnetic structures in DNS has been studied
in non-helical forced turbulence \citep{BKKMR11,BKR13,BGJKR14,WLBKR13,WLBKR16}
and in turbulent convection \citep{KBKMR12,KBKKR16}
with imposed weak horizontal or vertical magnetic fields.
However, NEMPI seems to work only when the magnetic field is
not too strong (magnetic energy density is less than the turbulent
kinetic energy density).

The formation of magnetic structures from a dynamo-generated field
has recently been studied for forced turbulence
\citep{MBKR14,Jabbari14,Jabbari15,Jabbari16} and
in turbulent convection \citep{MS16}.
In particular,
simulations by \cite{MBKR14} have shown that much stronger magnetic
structures can occur at the surface when the field is generated by a
large-scale $\alpha^2$ dynamo
in forced helical turbulence.
Subsequent work by \cite{Jabbari16} suggests that bipolar surface
structures are kept strongly concentrated by converging flow patterns
which, in turn, are produced by
a strong magnetic field through the Lorentz force.
This raises the question what kind of nonlinear interactions take place
when a turbulent dynamo operates in a density-stratified layer.
To investigate this problem in more detail, we study here
the dynamics of magnetic structures
both in DNS and MFS in similar parameter regimes.

The original work of \cite{MBKR14} employed a two-layer
system, where the turbulence is helical only in the lower part of
the system, while in the upper part it is nonhelical.
Such two-layer forced turbulence was also studied in spherical geometry
\citep{Jabbari15}.
They showed that in such a case, several bipolar structures form,
which later expand and create a band-like arrangement.
The two-layer system allowed us to separate the dynamo effect in the lower layer
from the effect of formation of intense bipolar structures in the upper layer.
The formation of flux concentrations from a dynamo-generated magnetic field in
spherical geometry was also investigated with MFS \citep{Jabbari13}.
In that paper,
NEMPI was considered as the mechanism creating flux concentrations.
Models similar to those of \cite{MBKR14} have also been studied by
\cite{Jabbari16}, who showed
that a two-layer setup is not necessary and that even a single layer
with helical forcing leads to formation of intense bipolar structures.
This simplifies matters, and such systems will therefore be studied
here in more detail before addressing associated MFS of corresponding
$\alpha^2$ dynamos.
In earlier work of \cite{MBKR14} and \cite{Jabbari16},
no conclusive explanation for the occurrence of bipolar structures
with a sharp boundary was presented.

We use both DNS and MFS to understand
the mechanism behind the nonlinear interactions resulting in the complex
dynamics of sharp bipolar spots.
One of the key features of such dynamics is the long lifetime of the
sharp bipolar spots that tend to persist several turbulent diffusion times.
It has been shown by \cite{Jabbari16} that the long-term existence
of these sharp magnetic structures is accompanied by the
phenomenon of turbulent magnetic reconnection in the vicinity
of current sheets between opposite magnetic polarities.
The measured reconnection rate was found to be nearly independent of
magnetic diffusivity and Lundquist number.

In this work, we study the formation and dynamics of sharp magnetic
structures both in one-layer DNS and in corresponding MFS.
We begin by discussing the model and the underlying equations both for
the DNS and the MFS (\Sec{TheModel}), and then present the results (\Sec{Results}),
where we focus on the comparison between DNS and MFS.
In the DNS, the dynamo is driven either directly by helically forced turbulence
or indirectly by nonhelically forced turbulence that becomes
helical through the combined effects of stratification and rotation,
as will be discussed at the end of \Sec{Results}.
We conclude in \Sec{Conclusions}.

\section{The model}
\label{TheModel}

We perform simulations in Cartesian coordinates
following \cite{Jabbari16}.
In our DNS, we study a one-layer model in which the forcing
is helical in the entire domain.
In the following we describe the details of both DNS and MFS.

\subsection{DNS equations}

First, we study an isothermally stratified layer
in DNS and solve the magnetohydrodynamic equations for the velocity $\UU$,
the magnetic vector potential $\AAA$, and the density $\rho$
in the presence of rotation $\Omega$,
\begin{equation}
{\partial\rho\over\partial t}=-\nab\cdot\rho\UU,
\end{equation}
\begin{equation}
\rho{\DD\UU\over\DD t}=\JJ\times\BB-\cs^2\nab\rho-2\OO\times\rho\UU
+\rho(\ff+\grav)+\nab\cdot(2\nu\rho\SSSS),\;\;
\end{equation}
\begin{equation}
{\partial\AAA\over\partial t}=\UU\times\BB+\eta\nabla^2\AAA,
\end{equation}
where the operator $\DD/\DD t=\partial/\partial t+\UU\cdot\nab$ is
the advective derivative,
$\OO=(-\sin\theta,0,\cos\theta)\Omega$ is the angular velocity
with $\theta$ being colatitude,
$\eta$ is the magnetic diffusivity,
$\grav=(0,0,-g)$ is the gravitational acceleration,
$\BB=\nab\times\AAA$ is the magnetic field,
$\JJ=\nab\times\BB/\mu_0$ is the current density,
${\sf S}_{ij}=\half(U_{i,j}+U_{j,i})-\onethird\delta_{ij}\nab\cdot\UU$
is the traceless rate of strain tensor (the commas denote
partial differentiation), $\nu$ is the kinematic viscosity,
$\cs$ is the isothermal sound speed, and $\mu_0$ is the vacuum permeability.
We adopt Cartesian coordinates $(x,y,z)$ and perform
isothermal simulations, so there is no possibility of convection.
Turbulence is produced by
the forcing function $\ff$ that consists of random, white-in-time,
plane waves with a certain average wavenumber $\kf$ \citep{B01,MBKR14}:
\begin{equation}
\ff(\xx,t)={\rm Re}\{N\fh(\kk,t)\exp[\ii\kk\cdot\xx+\ii\phi]\},
\end{equation}
where $\xx$ is the position vector.
We choose $N=f_0 \sqrt{\cs^3 |\kk|}$, where $f_0$ is a
nondimensional forcing amplitude.
At each timestep, we select randomly the phase
$-\pi<\phi\le\pi$ and the wavevector $\kk$
from many possible discrete wavevectors
in a certain range around a given forcing wavenumber, $\kf$.
Hence $\ff(t)$ is a stochastic process that is white-in-time and
is integrated by using the Euler--Maruyama scheme \citep{Hig01}.
The Fourier amplitudes,
\begin{equation}
\fh({\kk})=\RRRR\cdot\fh({\kk})^{\rm(nohel)}\quad\mbox{with}\quad
{\sf R}_{ij}={\delta_{ij}-\ii\sigma\epsilon_{ijk}\hat{k}
\over\sqrt{1+\sigma^2}},
\end{equation}
where the parameter $\sigma$ characterizes the fractional helicity of $\ff$, and
\begin{equation}
\fh({\kk})^{\rm(nohel)}=
\left(\kk\times\eee\right)/\sqrt{\kk^2-(\kk\cdot\eee)^2}
\label{nohel_forcing}
\end{equation}
is a non-helical forcing function. Here $\eee$ is an arbitrary unit vector
not aligned with $\kk$, $\hat{\kk}$ is the unit vecntor along $\kk$, and
$|\fh|^2=1$ \citep{B01}.
In most of the simulations, $\ff$ is maximally helical with
{\em positive} helicity, but we also consider cases without helicity.
The turbulent rms velocity is approximately
independent of $z$ with $\urms=\bra{\uu^2}^{1/2}\approx0.1\,\cs$.

We consider a cubic domain of size $L^3$ with $-L/2\leq x,y,z\leq L/2$
and define the base wavenumber as $k_1=2\pi/L$.
The density scale height is $H_\rho=\cs^2/g$,
where the value of $g$ is chosen such that
$k_1 H_\rho=1$, so the density contrast between
top and bottom is $\exp(2\pi)\approx535$.
In the following, we refer to $\kf/k_1$ as the scale separation ratio.

\subsection{MFS equations}

For the MFS, we consider the nonrotating case of a conducting isothermal gas
governed by the equations for the mean density $\meanrho$, the mean
(large-scale) velocity $\meanUU$, the mean vector potential $\meanAA$,
so that the mean magnetic field is given by $\meanBB=\nab\times\meanAA$.
Thus,
\begin{equation}
{\partial\meanrho\over\partial t}=-\nab\cdot\meanrho\meanUU,
\end{equation}
\begin{equation}
\meanrho{\meanD\,\meanUU\over\DD t}=\meanJJ\times\meanBB-\cs^2\nab\meanrho
+\meanrho\grav
+\nab\cdot(2\nuT\meanrho\meanSSSS),
\label{MNSE}
\end{equation}
\begin{equation}
{\partial\meanAA\over\partial t}
=\meanUU\times\meanBB+\alpha\meanBB+\etaT\nabla^2\meanAA,
\end{equation}
where $\alpha$ is given by \citep{Iro71}
\begin{equation}
\alpha={\alpha_0\over1+Q_\alpha\meanBB^2/\Beq^2},
\end{equation}
and $\meanD/\DD t=\partial/\partial t+\meanUU\cdot\nab$ is the advective
derivative with respect to the {\em mean} flow,
$\etaT$ and $\nuT$ are the total (sums of turbulent and microphysical)
magnetic diffusivity and kinematic viscosity, respectively,
$\alpha_0$ quantifies the kinematic $\alpha$ effect,
$Q_\alpha$ determines the strength of the $\alpha$ quenching,
$\meanJJ=\nab\times\meanBB/\mu_0$ is the mean current density,
$\meanSSSS$ is the traceless rate of strain tensor of the mean flow
with components $\meanSSS_{ij}=\half(\meanU_{i,j}+\meanU_{j,i})
-\onethird\delta_{ij}\nab\cdot\meanUU$,
and $\Beq=\sqrt{\mu_0\rho}\, \urms$ is the equipartition field strength.

\subsection{Boundary and initial conditions}

We adopt periodic boundary conditions in
the $x$ and $y$ directions and stress-free conditions at top and bottom
($z=\pm L/2$).
The magnetic field boundary conditions are perfect conductor at
the bottom and vertical field at the top.

In the MFS, we perform two-dimensional and three-dimensional simulations.
For the two-dimensional MFS we
adopt a squared-shaped Cartesian domain of size $L^2$ in the $x$
and $z$ directions, respectively.
Periodic boundary conditions are applied in the $x$ and $y$ directions
and perfectly conducting boundaries for the magnetic field
at the bottom ($z=-L/2$),
\EQ
A_x=A_y=A_{z,z}=0\quad\mbox{at $z=-L/2$ (bottom)},
\label{bot}
\EN
and vertical field conditions at the top ($z=L/2$)
\EQ
A_{x,z}=A_{y,z}=A_z=0\quad\mbox{at $z=L/2$ (top)}.
\label{top}
\EN
For the velocity field, both boundaries are assumed stress-free, i.e.,
\EQ
U_{x,z}=U_{y,z}=U_z=0
\quad\mbox{at $z=\pm L/2$}.
\EN
The same conditions apply to the MFS, but with $\AAA\to\meanAA$
and $\UU\to\meanUU$.
As initial conditions we adopt a hydrostatic equilibrium
with $\rho=\rho_0\exp(-z/H_\rho)$, where $\rho_0$ is a constant.
The initial magnetic field consists of weak gaussian-distributed noise.

\subsection{Parameters of the simulations}

Our units are chosen such that $\cs=g=\mu_0=1$.
In most of the calculations, we use $\kf/k_1=30$,
except in one case where we decrease it to $5$ to study
the effect of changing the scale separation ratio.
For the reference run, we use a Reynolds number
$\Rey\equiv\urms/\nu\kf$ of 100,
and a magnetic Prandtl number $\Pm=\nu/\eta$ of 0.5.
The magnetic Reynolds number, $\Rm\equiv\urms/\eta\kf$, is therefore
$\Rm=\Pm\Rey=50$.

Following earlier work by \cite{BCC09}, a more natural length scale
is given by the inverse wavenumber of the most slowly decaying mode,
which corresponds to a quarter wave and is given by
\EQ
\tilde{k}_1=\pi/2L.
\EN
With the boundary conditions \eq{bot} and \eq{top}, the most easily
excited solution corresponds to dynamo waves propagating in the positive
$z$ direction, as was found by \cite{BCC09}.
As in earlier work, a relevant timescale is the turbulent-diffusive time
given by
\EQ
\tautd=(\etaT\tilde{k}_1^2)^{-1}.
\EN
When comparing with earlier work of \cite{MBKR14} and \cite{Jabbari16},
we must remember that they defined the turbulent-diffusive time
$\tautd'=\tautd/16$ based on $k_1$.

The system is characterized by the following set of non-dimensional numbers:
the dynamo number and an analogous number characterizing $\urms$, i.e.,
\EQ
C_\alpha=\alpha_0/\etaT\tilde{k}_1,\quad
C_u=\urms/\etaT\tilde{k}_1,
\EN
as well as the turbulent magnetic Prandtl number and the Froude number,
\EQ
\Pmturb=\nuT/\etaT,\quad
\Fr=\urms\,\sqrt{\tilde{k}_1/g},
\EN
respectively.
In the MFS, it enters only indirectly through the definition of $\Beq$.
The value of $\Beq$ in the middle of the domain is $\Beqz=\Beq(z=0)$.
For the DNS of rotating turbulence, we also define the Coriolis number,
\EQ
\Co=2\Omega/\urms\kf.
\EN
All calculations have been performed with the {\sc Pencil Code}\footnote{
\url{https://github.com/pencil-code}}.
It uses sixth-order explicit finite differences in space
and a third-order accurate time-stepping method.
In the DNS, we adopt a numerical resolution of $256\times252\times256$
mesh points in the $x$, $y$, and $z$ directions in the Cartesian coordinate.
In the two- and three-dimensional MFS, we used $288^2$ or $288^3$
meshpoints, respectively.

\subsection{Simulation strategy}

As alluded to in the introduction, we want to study here a model that
is as simple as possible.
Before addressing the MFS, let us first consider the DNS.
The simpler one-layer model was already studied by \cite{Jabbari16};
see their Run~RM1zs.
In the following we focus on particular properties that are relevant for
our comparison with related MFS.

In the present work, we investigate the behavior of an $\alpha^2$ dynamo
and the formation of the structures with sharp boundaries.
We perform systematic parameter studies similar to \cite{Jabbari16}
to investigate the effect of changing magnetic Reynolds number
and scale separation on the structures.
Furthermore, in some runs we include the
Coriolis force in the momentum equation
to study the influence of rotation in our model.

\begin{table}\caption{
Summary of the DNS with helical forcing.
The reference run is shown in bold.
}\vspace{12pt}\centerline{\begin{tabular}{lrrcl}
Run & $\Rm$ & $\kf/k_1\!\!\!\!$ & $\Co$ & $\;\;\tilde{\lambda}$ \\
\hline
{\bf R1} & {\bf 50} & {\bf 30}& 0 & {\bf 0.013} \\
R2      &  130   & 30     & 0 &  0.002 \\
R3      &  260   & 30     & 0 &  0.001 \\
R4      &  50    & 5      & 0 &  0.028 \\
R5      &  50   & 30     & 0.3&  0.01~  \\
R6      &  50   & 30     & 0.7& 0.009 \\
R7      &  50   & 30     & 1.4& 0.005 \\
\label{Tab1}\end{tabular}}
\end{table}

\section{Results}
\label{Results}

In the following we start with DNS of helically forced turbulence,
compare with corresponding MFS, and finally study DNS with nonhelically
forced rotating turbulence.
In the latter case, the presence of rotation together with the density
stratification produce helicity and thus large-scale dynamo action.

\subsection{The one-layer model in DNS}

We begin with the helically forced case.
The parameters of our DNS with helically forced turbulence
are summarized in \Tab{Tab1}.
Here we also give a nondimensional estimate of the dynamo growth rate,
$\tilde{\lambda}=\lambda/\urms\kf$, where
$\lambda=\dd \ln B_{\rm rms}/\dd t$ is the instantaneous growth rate.
One can see that its value decreases with increasing
magnetic Reynolds number and with faster rotation, which
is consistent with the results of \cite{Jabbari16}; see their Fig.~2
and the discussion in their Section~3.

\subsubsection{Structure of the large-scale field}

\begin{figure}\begin{center}
\includegraphics[width=.99\columnwidth]{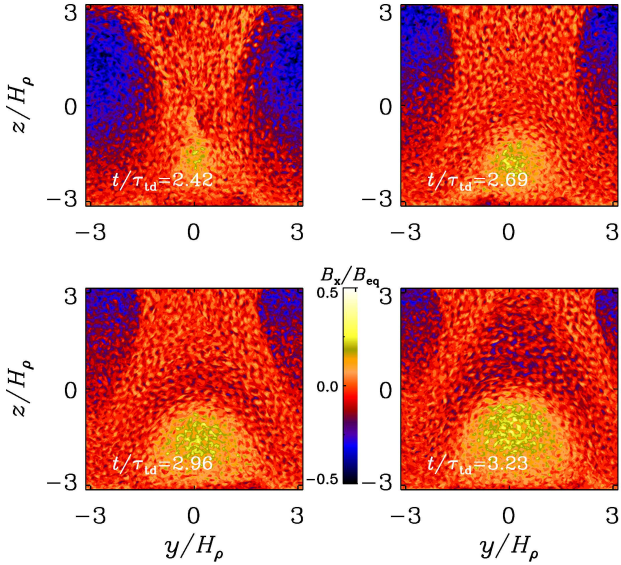}
\end{center}\caption{
DNS for $\Rm=50$ (Run~R1).
Time evolution of $B_{x}/\Beq$ in the $yz$ plane
through $x/H_\rho=\pi$.
}\label{pbx_yz_rpi}\end{figure}

\begin{figure}\begin{center}
\includegraphics[width=.99\columnwidth]{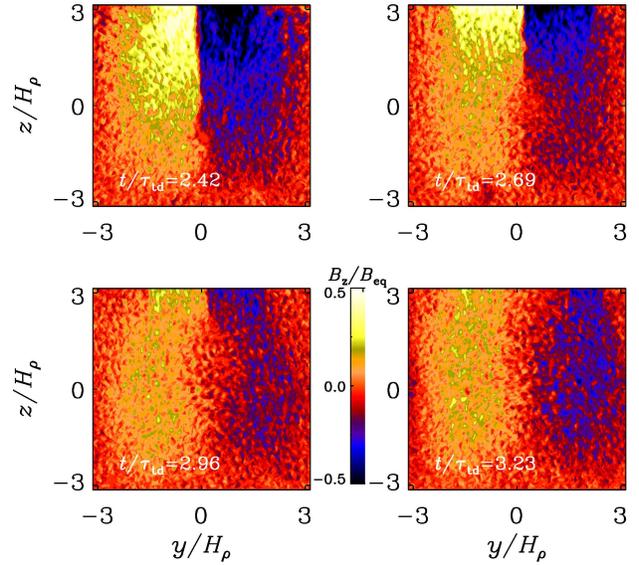}
\end{center}\caption{
DNS for $\Rm=50$ (Run~R1):
time evolution of $B_{z}/\Beq$ in the $yz$ plane
through $x/H_\rho=\pi$.
}\label{pbz_yz_rpi}\end{figure}

\begin{figure}\begin{center}
\includegraphics[width=.99\columnwidth]{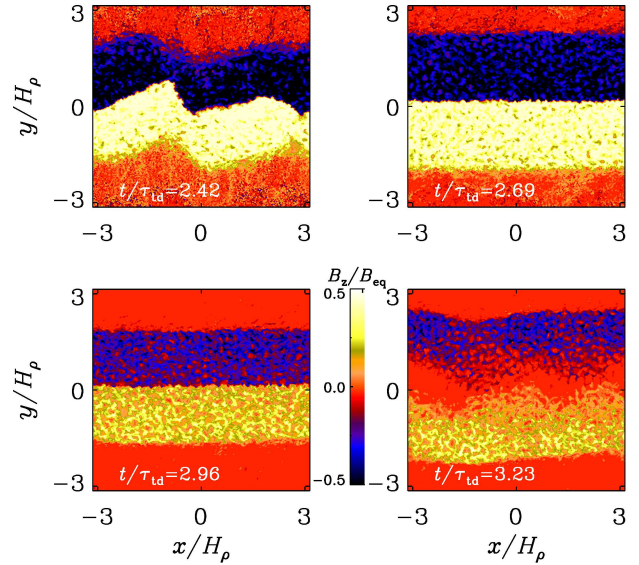}
\end{center}\caption{
DNS for $\Rm=50$:
time evolution of $B_z/\Beq$ in the $xy$ plane
through $z/H_\rho=\pi$ for Run~R1.
}\label{pbz_xy_rpi}\end{figure}

The dynamo-generated magnetic field is time-dependent.
This is caused by an underlying oscillatory $\alpha^2$-type dynamo
mechanism that has been seen in earlier DNS \citep{MTKB10,WBM11}.
It leads to a migratory dynamo wave from the lower perfect conductor
boundary toward the upper vacuum boundary \citep{BCC09,B16}.
Although the mean magnetic field of this dynamo has no $z$ component, it
develops one in the nonlinear stage, albeit with zero net vertical flux.
During certain times, the associated horizontal field locks into a state
where it is aligned with one of the two horizontal coordinate directions.
This alignment is a consequence of having adopted
a horizontally periodic domain.
In Run~R1, it points in the $x$ direction during the time
shown in \Fig{pbx_yz_rpi}, where we have selected an arbitrarily
chosen cross-section of $B_x$ in the $yz$ plane.
Clearly, the field is strongest in the deeper parts, $z/H_\rho\approx-2$,
and varies only little in the upper parts, $-1\leq z/H_\rho\leq\pi$.
By contrast, the vertical field $B_z$ is strongest in the upper parts,
$1\leq z/H_\rho\leq\pi$ (see \Fig{pbz_yz_rpi}), and develops sharp
structures
that are clearly seen at the top surface.
In particular, we see the formation of a sharp structure associated with
a {\sf Y}-point current sheet structure, similarly to that was shown in Figure~9
of \cite{Jabbari16}, where the associated reconnection phenomenology
was studied in detail.
The resulting surface structure
is shown in \Fig{pbz_xy_rpi},
where the formation of a current sheet and subsequent
reconnection of the magnetic field lines occur in the time interval
between $t/\tautd=2.4$ and $2.7$.

\subsubsection{Growth and evolution of the magnetic field}

\begin{figure}\begin{center}
\includegraphics[width=.9\columnwidth]{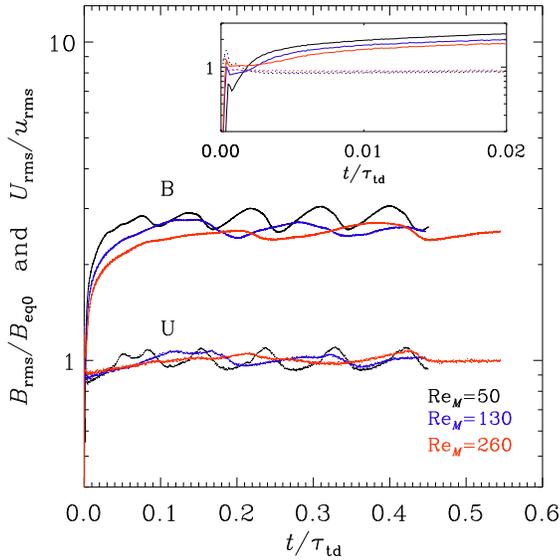}
\end{center}\caption{
DNS for $\Rm=50, 130, 260$:
$B_{\rm rms}/\Beqz$ (solid) and $U_{\rm rms}/\urms$ (dashed)
versus time.
Different colors represent different values of $\Rm$:
50 for Run~R1 (black), 130 for Run~R2 (blue), and 260 for Run~R3 (red).
The inset shows the early times.
}
\label{pubrms_comp}\end{figure}

In all of our simulations, the initial magnetic field
grows rapidly to become comparable to $\Beq$,
so no clear kinematic dynamo stage can be seen.
This has the advantage that these simulations reach quickly a nonlinear
statistically steady state.
On the other hand, for such strong magnetic fields NEMPI cannot be
observed in DNS.
Similar to our earlier work, we find that, in the nonlinear stage,
the amplitudes of the oscillations of the kinetic
and magnetic energy densities are small; see \Fig{pubrms_comp}.

\begin{figure}\begin{center}
\includegraphics[width=.8\columnwidth]{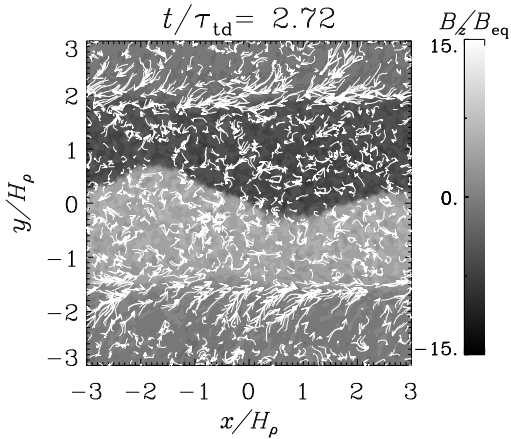}
\end{center}\caption{
Velocity vectors $(U_x,U_y)$ at the surface ($z/H_\rho=\pi$) for Run~R1
(DNS for $\Rm=50$) superimposed on a gray-scale representation of
$B_{z}/\Beq(z)$ at $t/\tautd=2.46$.
}\label{pbz_one_vec_rpi}\end{figure}

\begin{figure}\begin{center}
\includegraphics[width=.98\columnwidth]{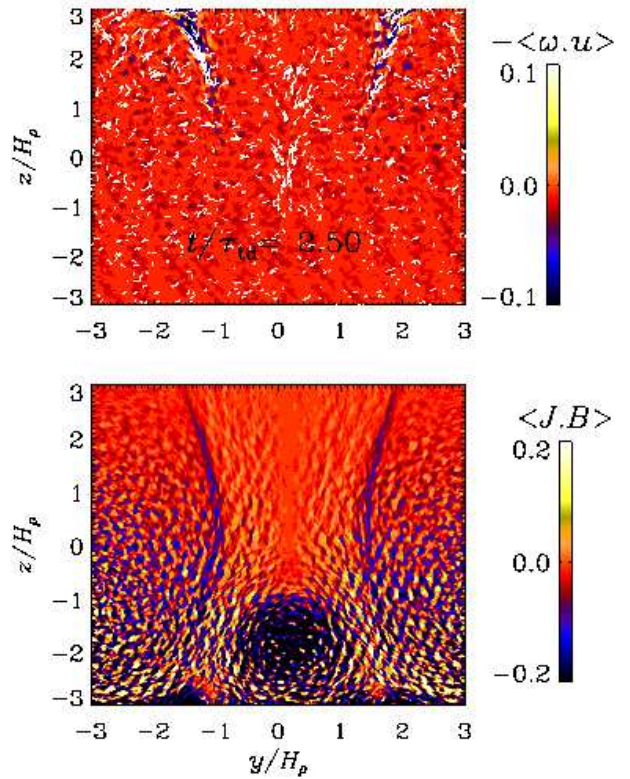}
\end{center}\caption{
DNS for $\Rm=50$ (Run~R1):
negative mean kinetic and positive current helicity densities
averaged along the $x$ direction.
}\label{phel}\end{figure}

\subsubsection{Horizontal flows along magnetic boundaries}

We tend to find systematic horizontal flows along magnetic boundaries.
This is demonstrated in \Fig{pbz_one_vec_rpi}, where we show
horizontal flow vectors together with a gray-scale representation
of the magnetic field.
These horizontal flows are in opposite directions on both sides.
If those horizontal flows are associated with systematic vertical flows,
they would imply a systematic helicity.
In \Fig{phel} we plot the mean kinetic helicity, $\bra{\oo\cdot\uu}$,
where $\oo=\nab\times\uu$ is vorticity,
and the mean current helicity, $\bra{\JJ\cdot\BB}$.
These quantities are averaged along the $x$ direction.
We see that $\bra{\oo\cdot\uu}$ shows pronounced {\em positive} extrema
near the regions which turn out to coincide with downdrafts.
The sign of the kinetic helicity is surprising
because, if the reconnection regions are
associated with significant downflows,
the corresponding kinetic helicity should be
negative, because
$u_z<0$ in the downflows would coincides with right-handed
(anticlockwise) swirling motions with $\omega_z>0$.

\begin{figure}\begin{center}
\includegraphics[width=.99\columnwidth]{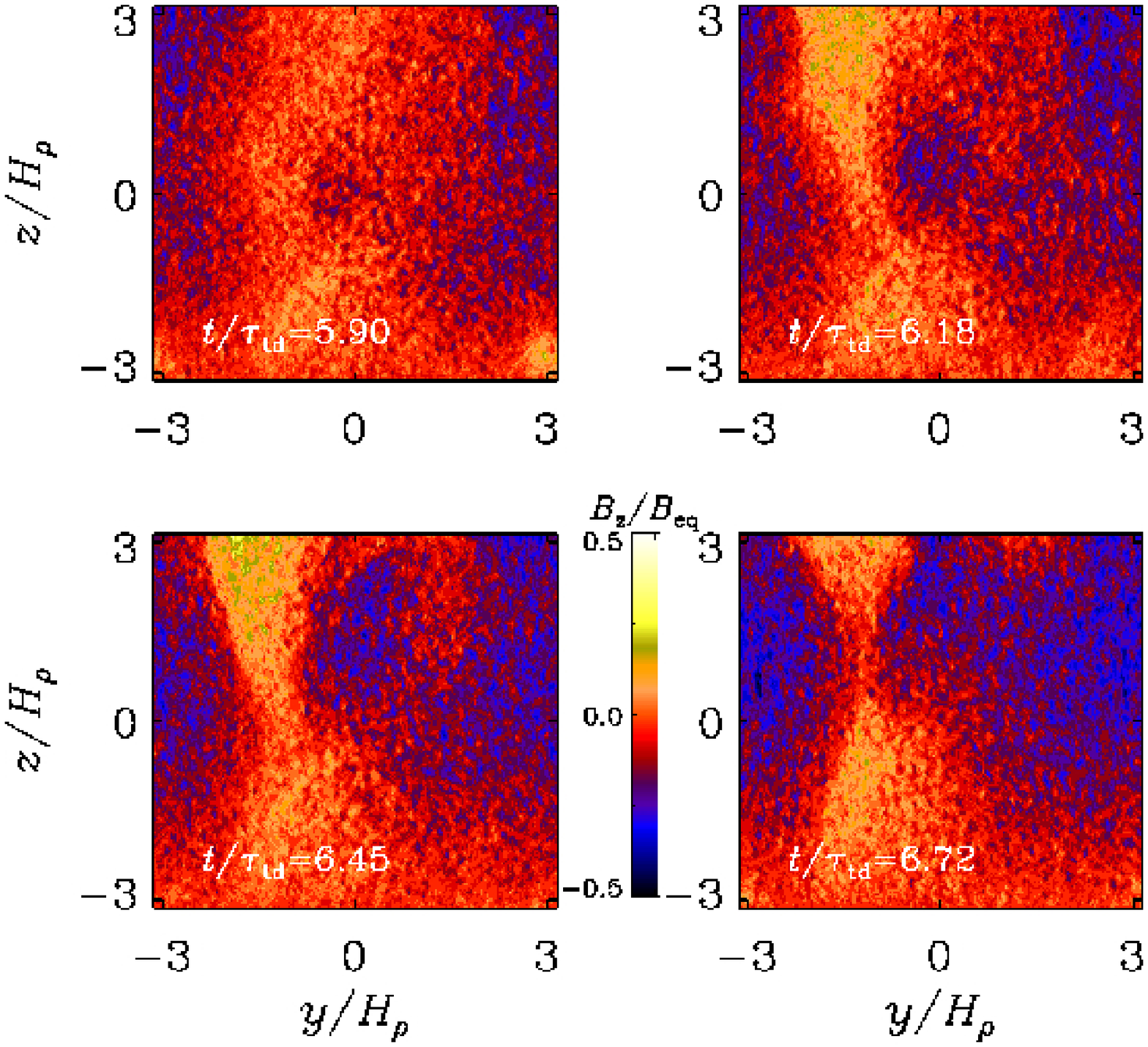}
\end{center}\caption{
DNS for $\Rm=260$ (Run~R3):
time evolution of $B_{z}/\Beq$ in the $yz$ plane
through $x/H_\rho=\pi$ for the fully helical run.
}\label{pbz_yz_rpi_260}\end{figure}

\begin{figure}\begin{center}
\includegraphics[width=.99\columnwidth]{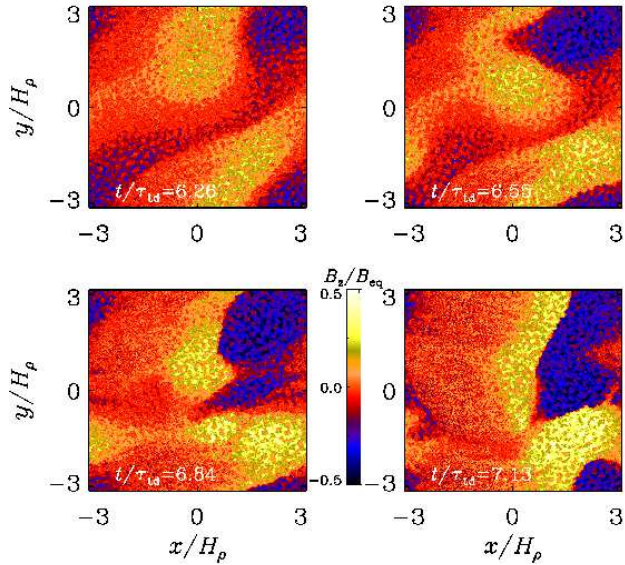}
\end{center}\caption{
DNS for $\Rm=260$ (Run~R3):
time evolution of $B_z/\Beq$ in the $xy$ plane
at the top surface for the fully helical run.
}\label{pbz_xy_rpi_260}\end{figure}

\subsubsection{Higher Reynolds numbers
and smaller scale separation}

At larger values of $\Rm$, the surface appearance of the field
becomes more fragmented.
This might be interesting in view of sunspot formation, since
active regions appear to be more isolated than what one expects
from a diffusive large-scale magnetic field.

Here we study the effects of varying the magnetic Reynolds number
on the formation of structures in the one-layer model.
In \Figs{pbz_yz_rpi_260}{pbz_xy_rpi_260}, we
present the results for $\Rm=260$.
We recall that $\Pm$ was kept constant ($\Pm=0.5$),
which implies that $\Rey$ varies from 100 to up to 500 in these simulations.
One can see from \Fig{pubrms_comp} that increasing the value of $\Rm$
leads to a decrease in the amplitude of the nonlinear oscillations.

Next, we study the formation of sharp structures as seen in \Fig{pbz_xy_rpi}.
For this purpose, we use $x$-averaged data because
the resulting structure is independent of $x$ at the time the structure has developed.
This does not apply to the run with the highest $\Rm$ (Run~R3) where the structure
does depend on $x$; see \Fig{pbz_xy_rpi_260}.

To confirm that
in our one-layer model the formation of the bipolar magnetic structures
is independent of the value of the scale separation ratio, we now
consider the case with $\kf/k_1=5$ (Run~R4).
The main difference relative to our reference model is that the structures
move faster and are more irregular.
They also form at later times relative to the reference run
with larger scale separation.

\subsubsection{Effect of rotation}

In this section we consider DNS of helically forced rotating turbulence.
We see from \Tab{Tab1} that an increase in the
rotation rate leads to a decrease in the growth rate of the dynamo when
$\Co$ is of the order of unity (cf.\ Runs~R5 and R6).
However, even for $\Co=1.4$ (Run~R7), sharp structures can still form,
as was already emphasized in earlier work \citep{Jabbari16},

Owing to the presence of stratification, rotation leads to the additional
production of kinetic helicity.
Once rotation is fast enough, the resulting helicity will lead to
an $\alpha$ effect that can be supercritical for dynamo action.
We return to this at the end of the paper.

\subsubsection{Energy spectra}

\begin{figure}\begin{center}
\includegraphics[width=.98\columnwidth]{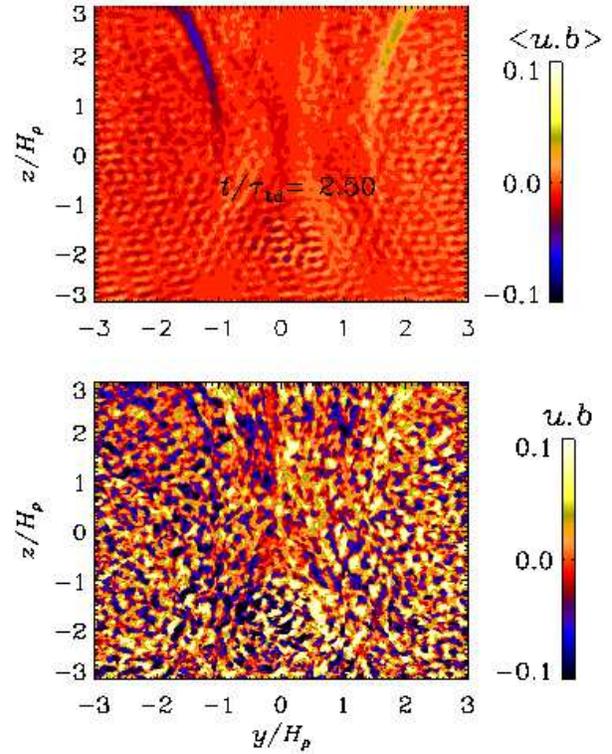}
\end{center}\caption{
DNS for $\Rm=50$ (Run~R1):
cross helicity density (upper panel) averaged along the $x$ direction.
The lower panel shows the cross helicity at $x/H_\rho=0.5$.
}
\label{phel_cross}\end{figure}

Similar to earlier findings in strongly stratified layers
\citep{BGJKR14}, there is a dramatic build-up of power
at the largest scales of the domain.
This was tentatively associated with inverse cascade-like
behavior that could be related with the production of
cross helicity, $\bra{\uu\cdot\BB}$, due to the presence
of a mean magnetic field parallel or antiparallel to gravity.
Indeed, strong cross helicity is also present in our current model,
but its sign changes across the surface, because the mean
vertical field changes; see \Fig{phel_cross}.

\begin{figure*}\begin{center}
\includegraphics[width=.65\columnwidth]{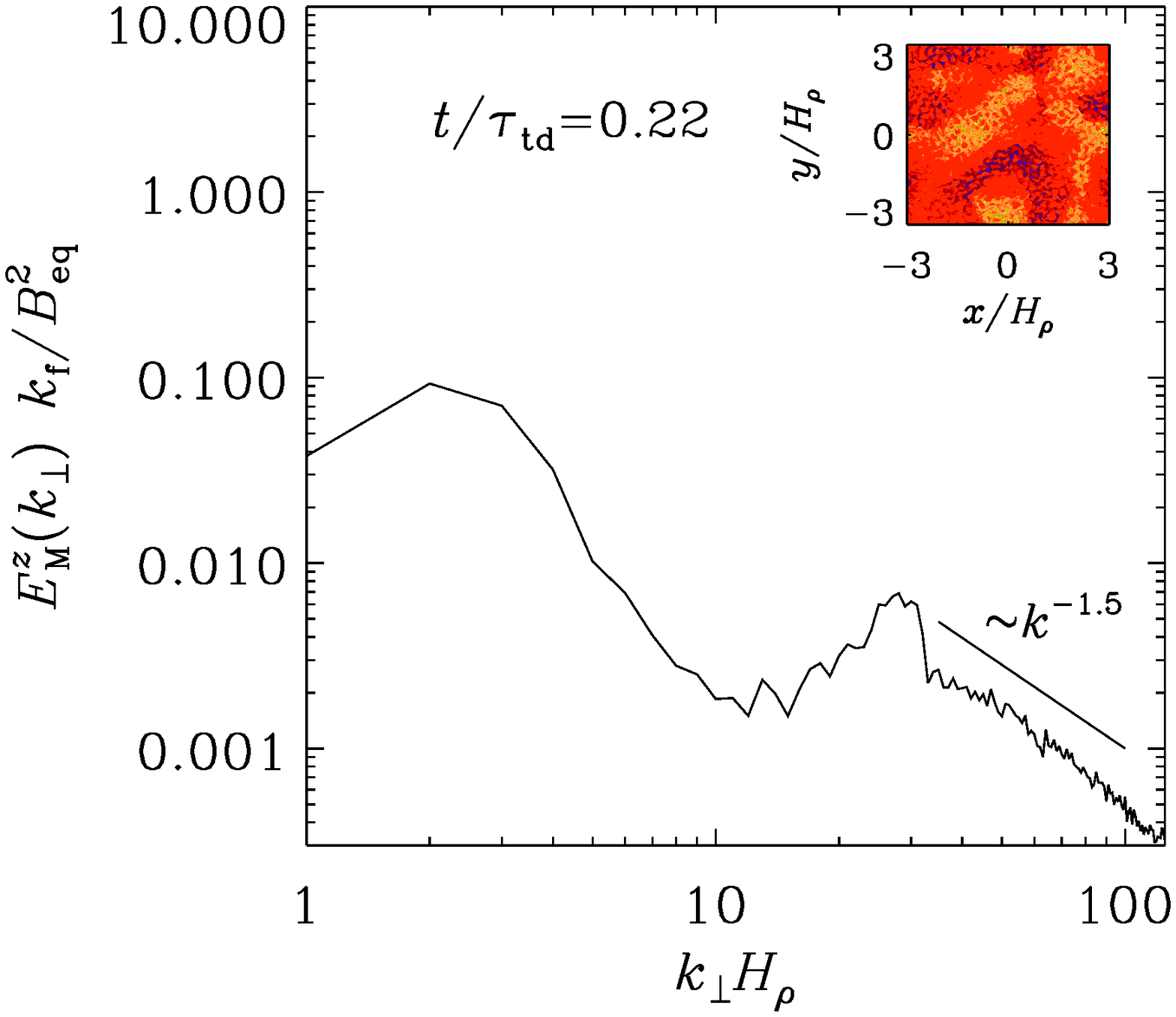}
\includegraphics[width=.65\columnwidth]{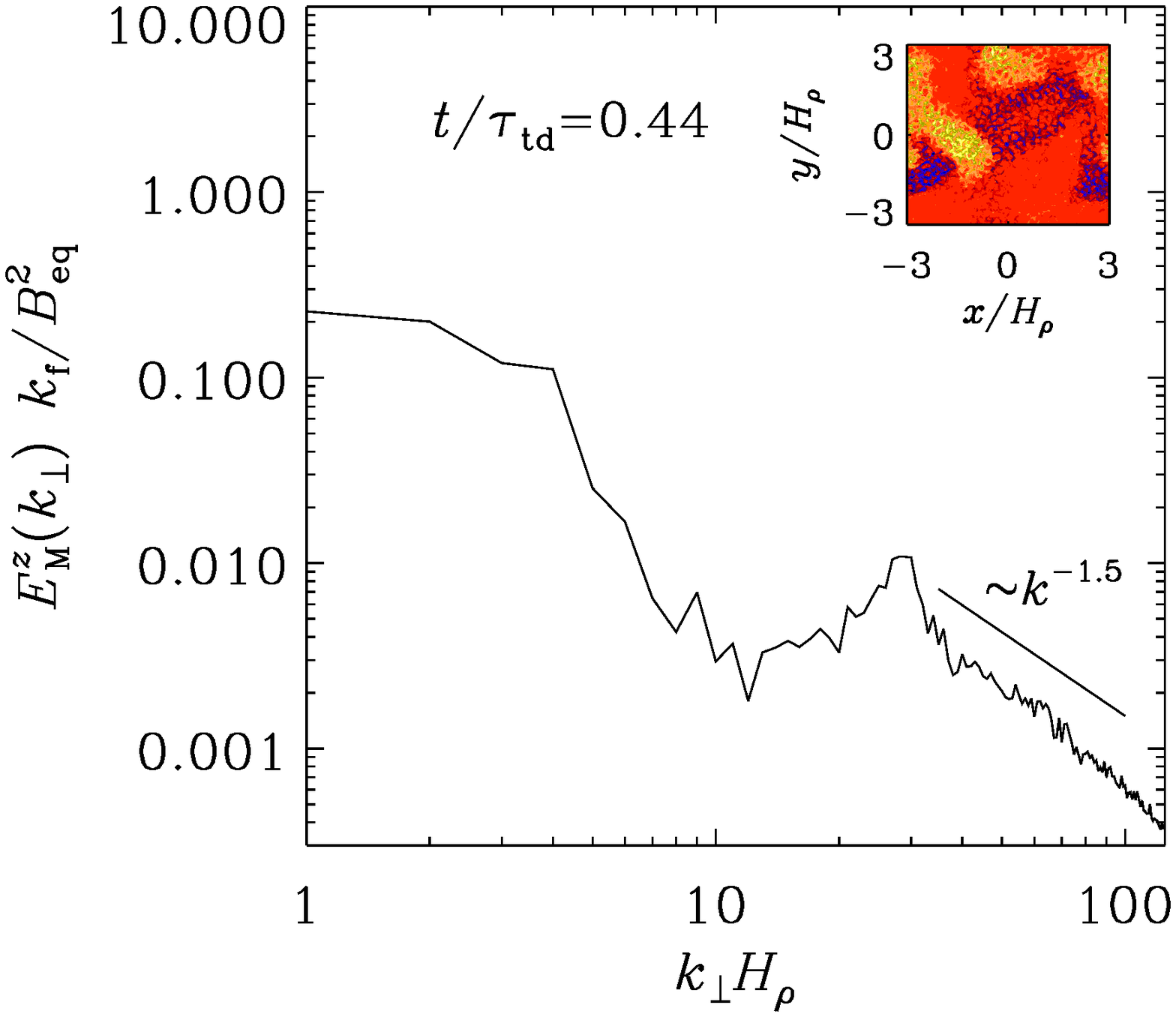}
\includegraphics[width=.65\columnwidth]{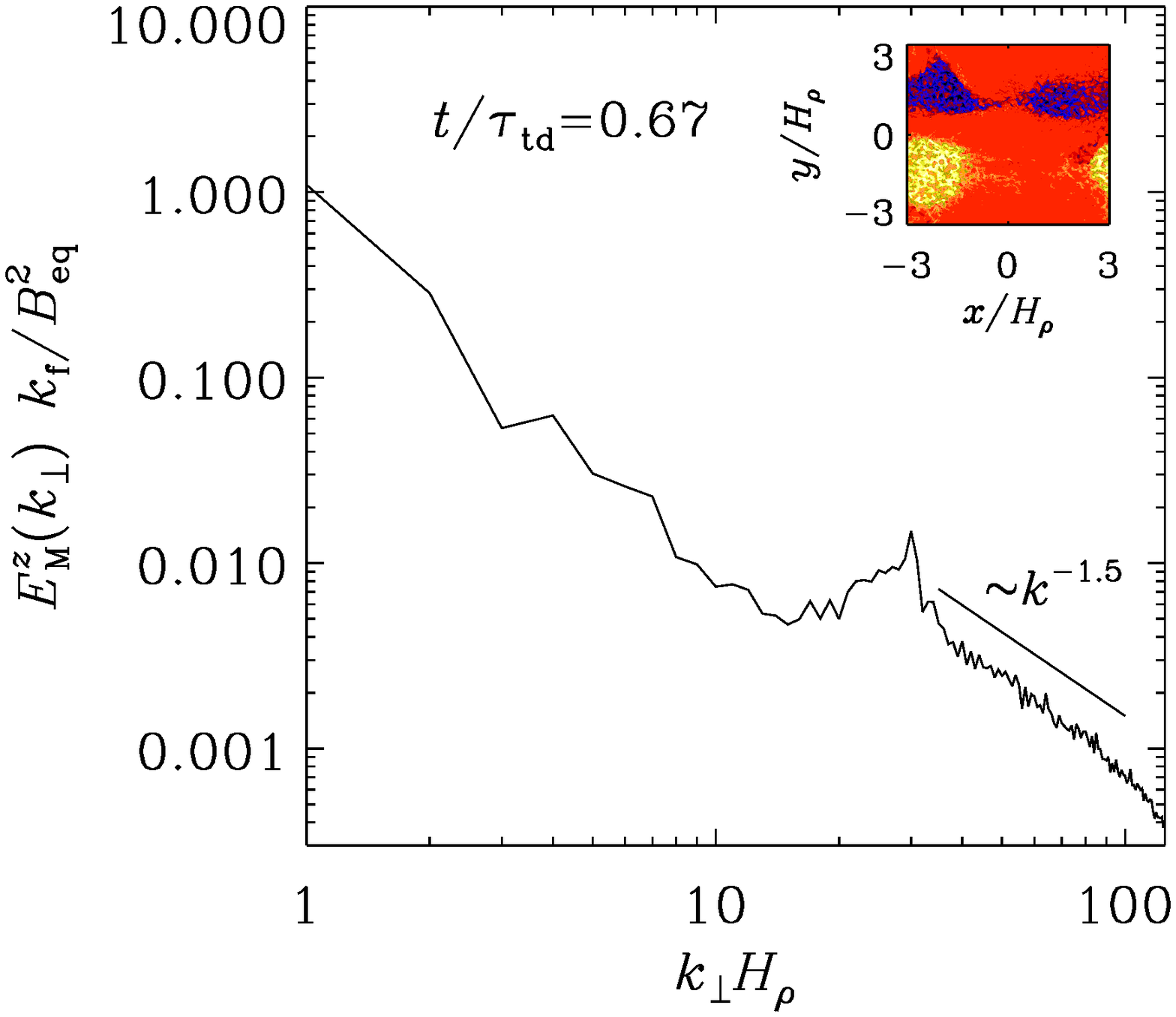}\\
\includegraphics[width=.65\columnwidth]{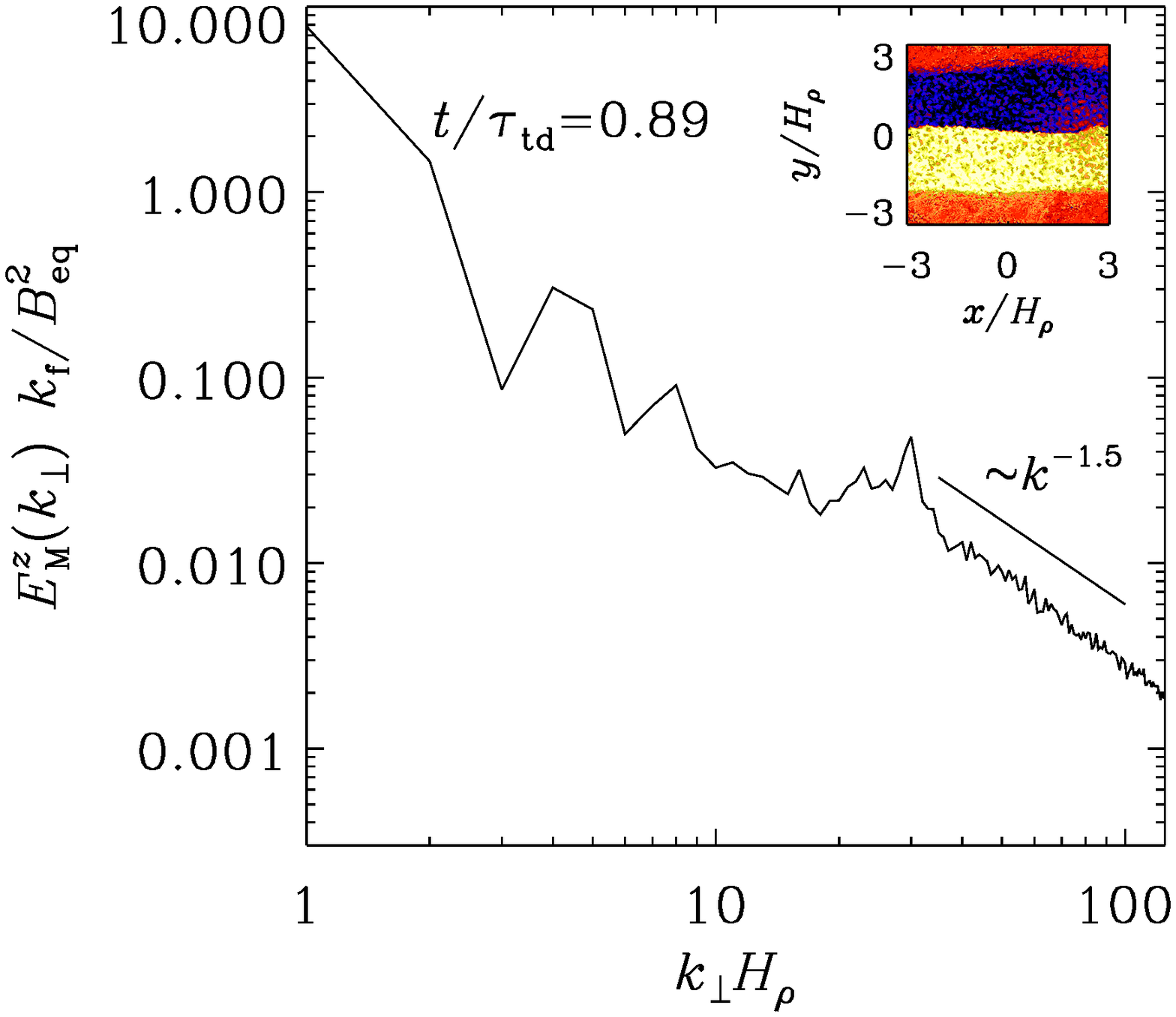}
\includegraphics[width=.65\columnwidth]{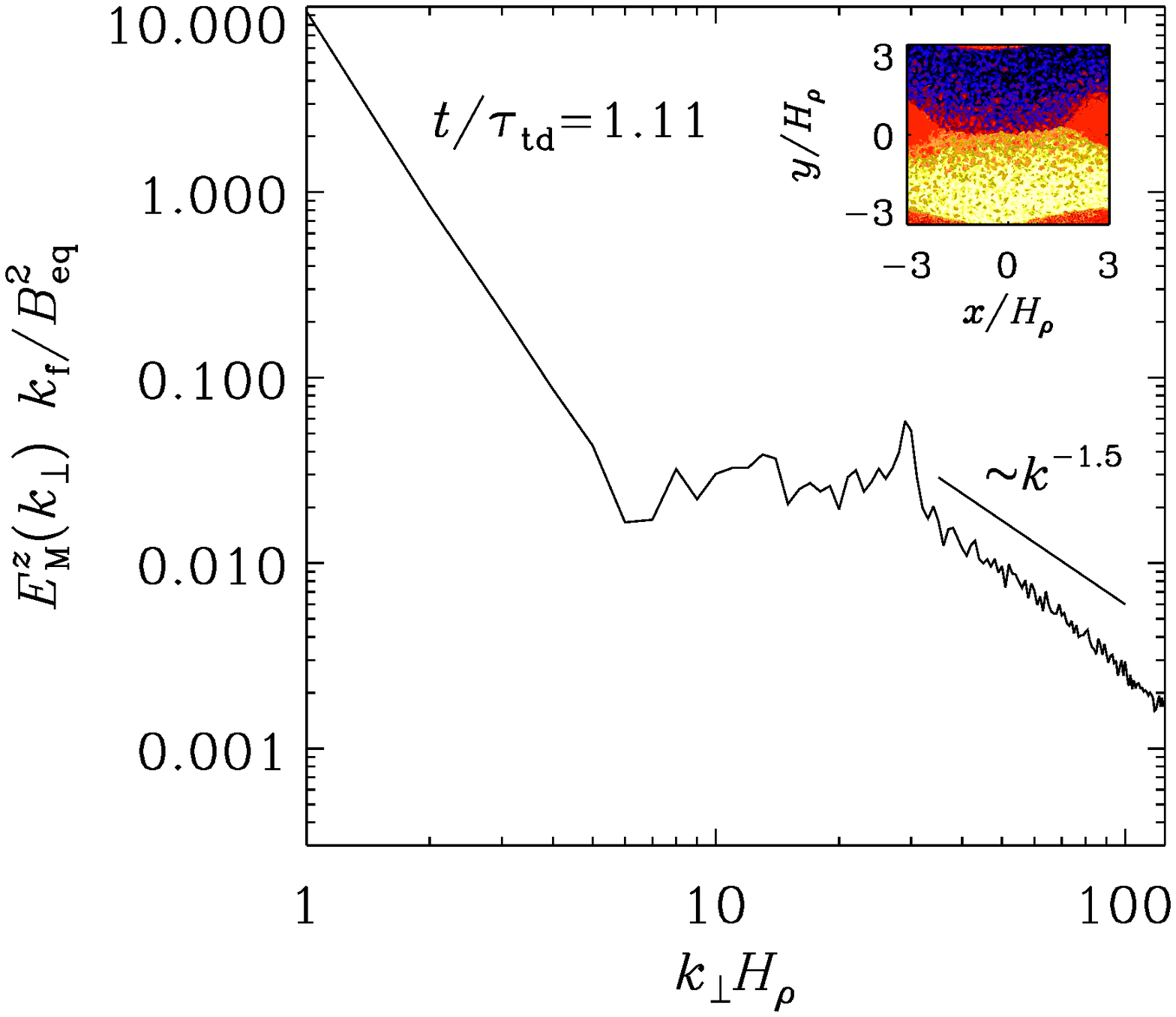}
\includegraphics[width=.65\columnwidth]{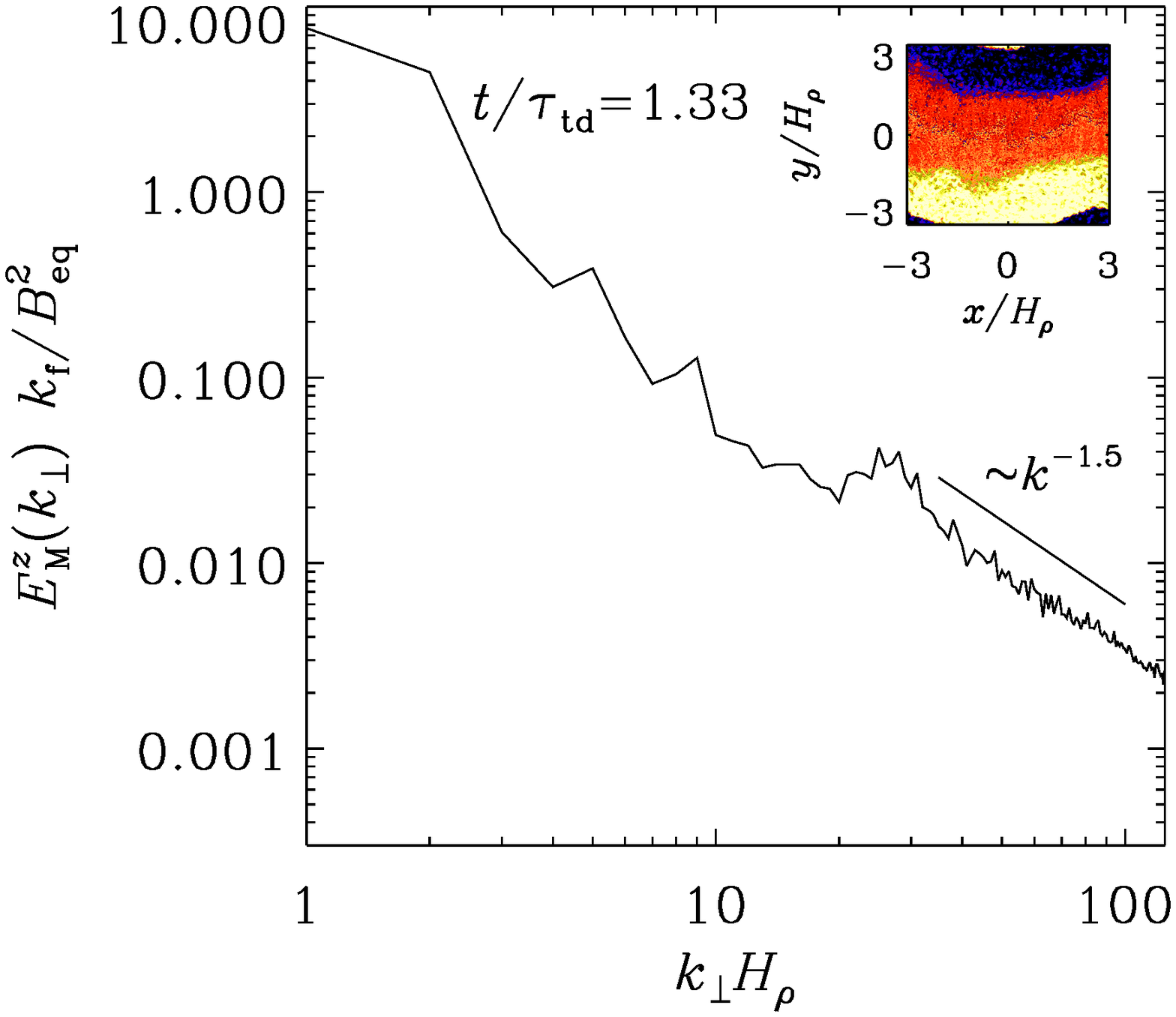}
\end{center}\caption{
DNS for $\Rm=50$ (Run~R3):
time evolution of $E_{\rm M}^z(k_\perp)$ normalized by $\Beq^2/\kf$.
The insets show the vertical field at the surface.
}
\label{power_series}\end{figure*}

The dramatic build-up of power at the largest scales is best
demonstrated by plotting horizontal power spectra of
$B_z(x,y)$ taken at the top of the domain during
the formation of the structures; see \Fig{power_series}.
These spectra denote the energy in wavenumber shells of radius
$k_\perp=(k_x^2+k_y^2)^{1/2}$, and are normalized such that
$\int E_{\rm M}^z(k_\perp) \, \dd k_\perp=\bra{B_z^2}/2$.
Note that the ratio between the energy injection wavenumber
and the wavenumber of the peak of the spectrum is equal to
the scale separation ratio, $\kf/k_1=30$.

\subsection{The dynamo in MFS}

To understand the origin and nonlinear dynamics of the sharp structures
found in the present DNS in the one-layer model and in the two-layer models
of \cite{MBKR14} and \cite{Jabbari16},
we start with a simple two-dimensional mean-field one-layer
model with an algebraically quenched $\alpha$ effect and
feedback from the self-generated large-scale magnetic field.
In Sect.~\ref{3DMFS}, we also consider a three-dimensional mean-field one-layer
model.
For simplicity, we have ignored here the dynamical
nonlinearity caused by the evolution of the magnetic helicity.
Furthermore, we have ignored algebraic quenching of the turbulent diffusivity.
Since the dynamo growth is very rapid in DNS, we begin by neglecting in
our mean-field model the effects of NEMPI, i.e., the effects of turbulence
on the mean effective magnetic pressure.
We do, however, include NEMPI in some of the 3D MFS; see Sect.~\ref{MFS-NEMPI}.

\subsubsection{Magnetic field evolution in MFS}

Mean-field dynamo action begins when $C_\alpha>C_\alpha^{\rm crit}$,
i.e., when $C_\alpha$ exceeds a critical value.
Using 2D MFS we find that the dynamo threshold $C_\alpha^{\rm crit}\approx2.55$;
which agrees with the analytic value of $2.550650$ \citep{B16}.
The cycles appear particularly pronounced in the rms velocity
of the mean flow $\meanUU$, which excludes the small-scale
flow that is implicitly present in the MFS in that it provides
the turbulent diffusion.
If this component is included, then the resulting total
rms velocities show much less variation with the cycle.
The cycle frequency for the marginally exited state is
$\omega_{\rm cyc}=2\pi/P_{\rm cyc}\approx1.43\tautd^{-1}$,
where $P_{\rm cyc}\approx4.4\tautd$ is the cycle period.
Again, this is in good agreement with the analytic value of
$1.429692$ \citep{B16}.

\begin{figure}\begin{center}
\includegraphics[width=.9\columnwidth]{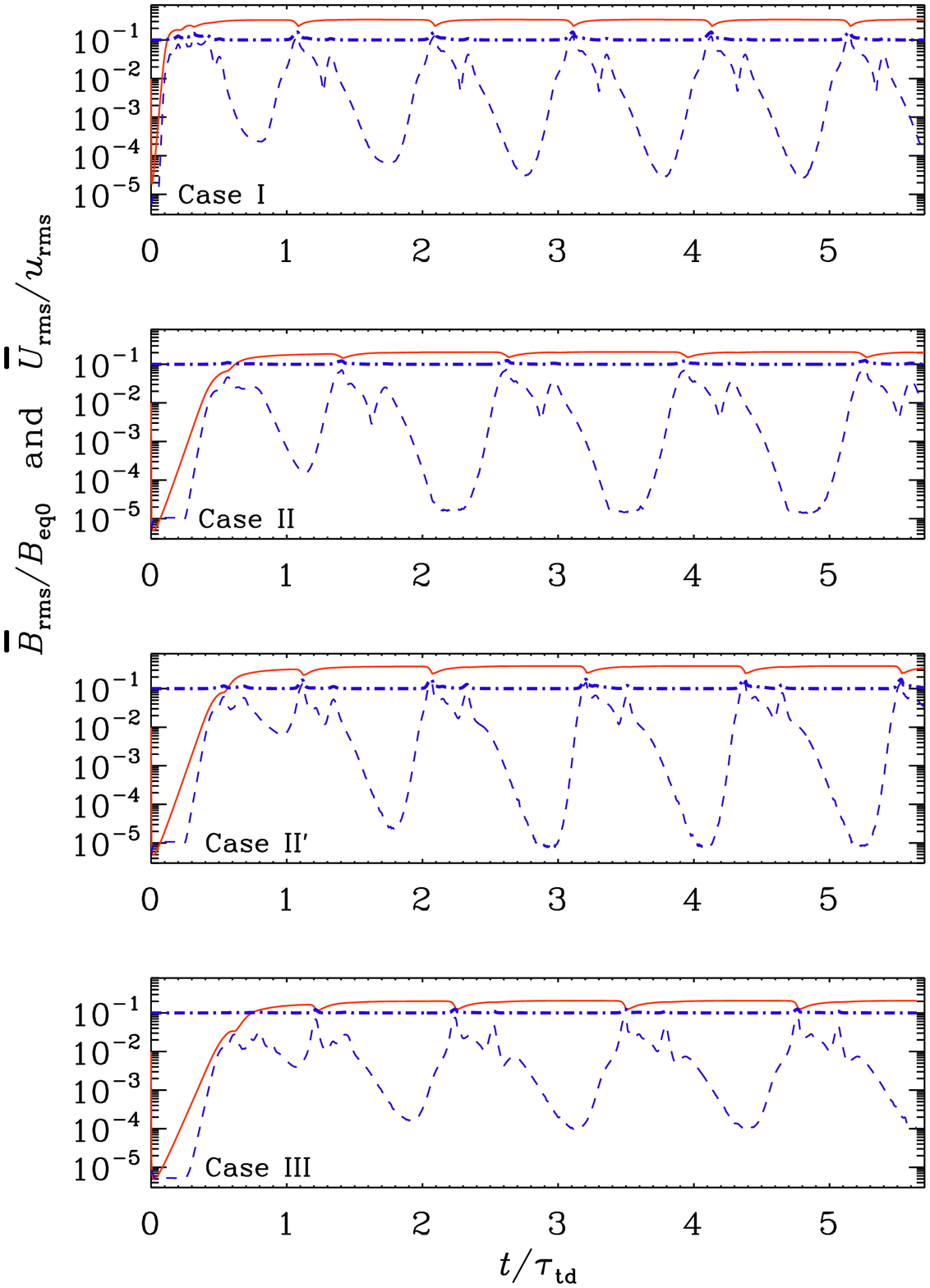}
\end{center}\caption{
2D MFS:
$\meanBrms/\Beqz$ (solid red) and $\meanUrms/\urms$ (dashed blue) vs.\ time.
The fat dash-dotted line gives the total rms velocity,
which includes the contribution from $\urms$.
}\label{pubrms}\end{figure}

In the following, we fix $\Pmturb=1$ and $\Fr=0.05$.
These values agree with those adopted
in the DNS of \cite{MBKR14} and \cite{Jabbari16}.
The values of $C_\alpha$ and $C_u$ are harder to estimate from the DNS.
We consider several cases that are listed in \Tab{TSum}.
The time evolutions of
$\meanUrms/\urms$ and $\meanBrms/\Beqz$
are shown in \Fig{pubrms}.
In all cases, there are long-term oscillations with
a period of approximately the turbulent-diffusive time.
The cycle frequencies are listed in \Tab{TSum} and compared with the
marginally excited (linear) case,
whose cycle frequency is less than those in all the nonlinear cases.
As usual, the cycle frequency is determined from any of the nonvanishing
components of $\meanBB$ and thus not from $\meanBrms$.
The normalized frequencies increase slightly with increasing
value of $C_u$; cf.\ Runs III-V in \Tab{TSum}.
The nonlinear oscillations are particularly pronounced in the mean flow.
When the mean flow reaches a maximum, the mean magnetic field
strength decreases slightly.
It is seen in \Fig{pubrms} that the minima of the rms value of the
mean velocity during the cycle are much deeper than those
of the mean magnetic field.

\begin{table}\caption{
Summary of MFS; `marg' refers to the marginally excited case,
which is independent of the values of $C_u$ and $Q_\alpha$.
}\vspace{12pt}\centerline{\begin{tabular}{cccccc}
Case & $C_\alpha$ & $C_u$ & $Q_\alpha$ &
$\omega_{\rm cyc}\,\tautd$ & $\meanBrms/\Beqz$ \\
\hline
 I & 20 & 40 &  1  & 3.08 & 0.32 \\ 
 I3D & 20 & 40 &  1  & 6.23 & 0.33 \\ 
 II & 10 & 40 &  1  & 2.38 & 0.20 \\ 
 II' & 10 & 40 & 0.3 & 2.68 & 0.37 \\ 
 II3D & 10 & 40 &  1  & 4.5 & 0.20 \\ 
 II3D-rot & 10 & 40 &  1  & 1.9 & 0.21 \\ 
 III & 10 & 80 &  1  & 3.11 & 0.19 \\ 
 IV & 10 &160 &  1  & 3.62 & 0.19 \\ 
 V & 10 &320 &  1  & 4.04 & 0.19 \\ 
VI3D & 5 &40 & 1  & 1.8 & 0.11 \\ 
VI3D-NEMPI & 5 & 40 &  1  & 1.53 & 0.11 \\ 
marg&2.55& ---& --- & 1.43 &  --- \\ 
\label{TSum}\end{tabular}}\end{table}

\begin{figure}\begin{center}
\includegraphics[width=\columnwidth]{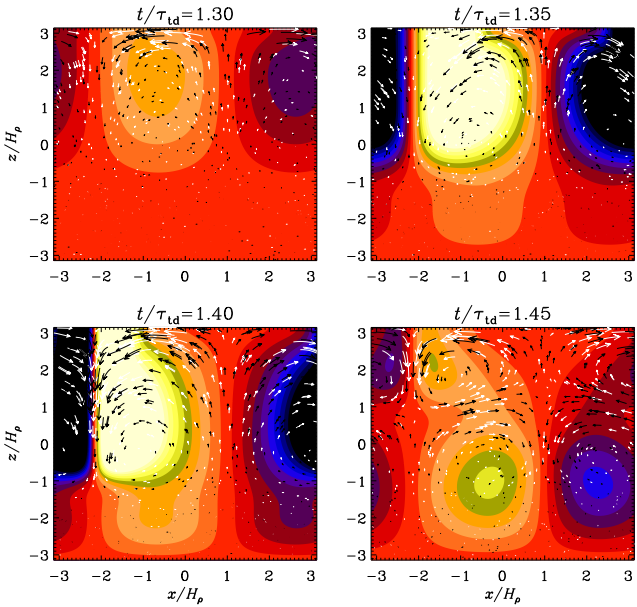}
\end{center}\caption{
2D MFS:
snapshots of $\meanB_z$ (color-coded) together
with vectors of $\meanUU$ in the $xz$ plane for case~II
during the time when the mean flow reaches its maximum.
Note the sharp structure at $t/\tautd=1.40$.
}\label{288d}\end{figure}

In \Fig{288d} we show four snapshots of $\meanB_z$ together
with vectors of $\meanUU$ in the $xz$ plane during the
time when the mean flow reaches its maximum.
The maxima in $\meanUrms$ correspond to times when
a strong downflow develops.
As is evident from \Fig{288d}, these flows push fields of
opposite sign together and form a current sheet in the
uppermost layer, which then leads to a downflow.
This destroys most of the field through turbulent diffusion
(or turbulent reconnection), but it is soon being replenished
by dynamo action from the deeper layers.

\begin{figure}\begin{center}
\includegraphics[width=\columnwidth]{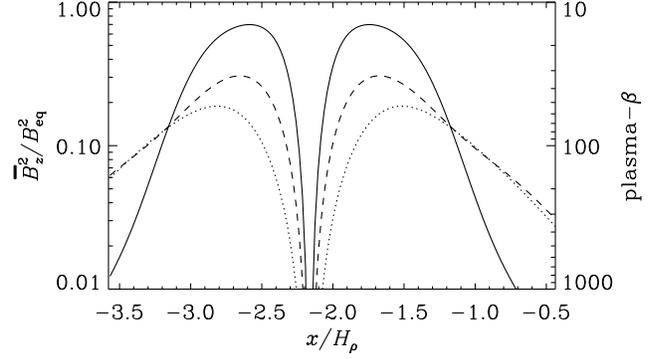}
\end{center}\caption{
2D MFS: $\meanBB^2/\Beq^2$ (left axis) and plasma-$\beta$ (right axis) as a
function of $x/H_\rho$ for $t/\tautd=1.40$ through $z/H_\rho=3$ (solid line),
$2$ (dashed line), and $1$ (dotted line) for case~II.
The $x$ axis has been extended periodically to smaller values.
}\label{pbeta}\end{figure}

\begin{figure}\begin{center}
\includegraphics[width=\columnwidth]{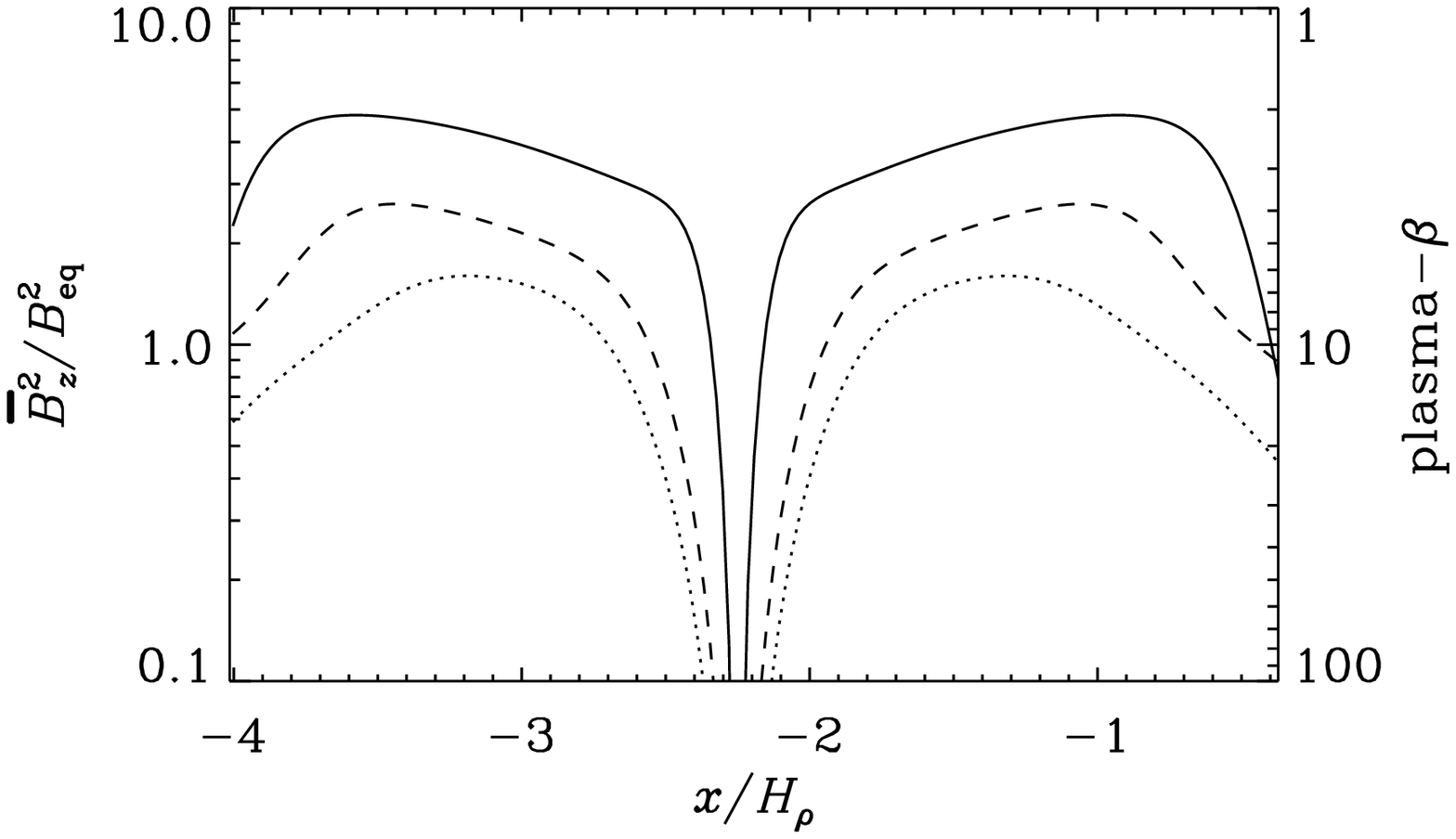}
\end{center}\caption{
2D MFS: similar to \Fig{pbeta}, but for case~II'.
}\label{pbeta_288f}\end{figure}

In our model, the $\alpha$ quenching with $Q_\alpha=1$ limits the
field strength to values just below the local equipartition value.
This is clear from \Fig{pbeta}, where we show
$\meanB_z^2/\Beq^2$ versus $x/H_\rho$ for $z/H_\rho=3$, $2$, and $1$.
The plasma-$\beta$ of the vertical field defined as
$\mu_0\meanrho\cs^2/\meanB_z^2$, reaches minimum values of around 10.
However, in the modified model (case~II') with $Q_\alpha=0.3$, the
minimum plasma-$\beta$ reaches values of about 2; see \Fig{pbeta_288f}.
In fact, even smaller values of $Q_\alpha$ down to 0.11 have been found
in rotating convection using the test-field method; see \cite{KRBKK14}.

\begin{figure}\begin{center}
\includegraphics[width=.99\columnwidth]{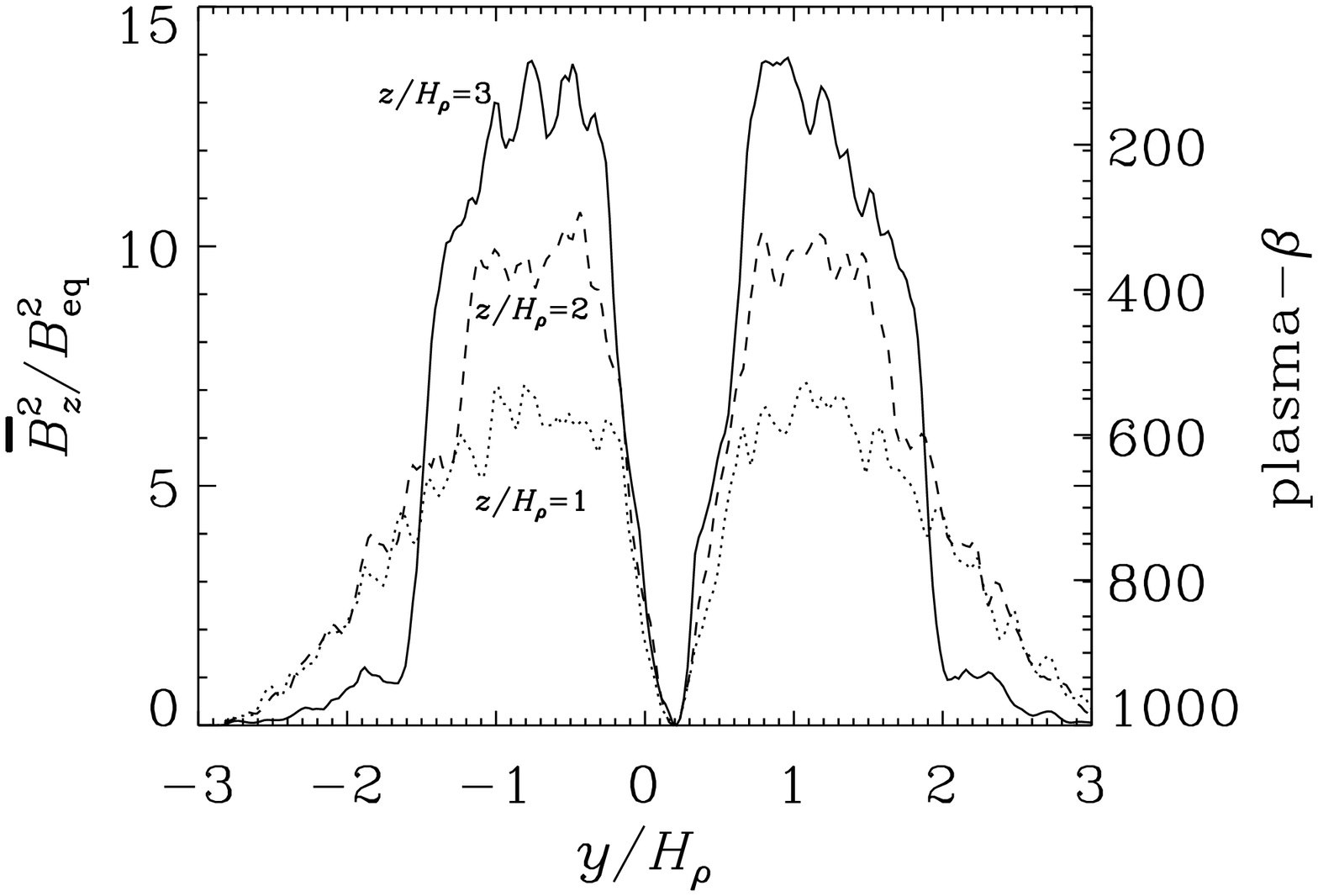}
\end{center}\caption{
DNS for $\Rm=50$ (Run~R1):
$\meanBB^2/\Beq^2$ as a function of $y/H_\rho$ for $t/\tautd=2.50$ through
$z/H_\rho=3$ (solid line), $2$ (dashed line), and $1$ (dotted line).
}\label{pbz2y}\end{figure}

\subsubsection{Height dependence of sharp structures}

The sharp structures are particularly pronounced in the upper,
low density regions.
We have seen this already in \Figs{pbeta}{pbeta_288f}, where we show
$\meanB_z^2/\Beq^2$ for three values of $z/H_\rho$ as a function of $y$.
The same behavior is also seen in the DNS in \Fig{pbz2y}, where we plot
the same quantity, which is now shown as an $x$-averaged quantity at the
time $t/\tautd=2.50$, when the magnetic structures are sharpest.
Comparing with \Figs{pbeta}{pbeta_288f}, we also see that the
larger values of $\meanB_z^2/\Beq^2$ in the DNS are best matched for
a smaller value of $Q_\alpha$.

\begin{figure}\begin{center}
\includegraphics[width=\columnwidth]{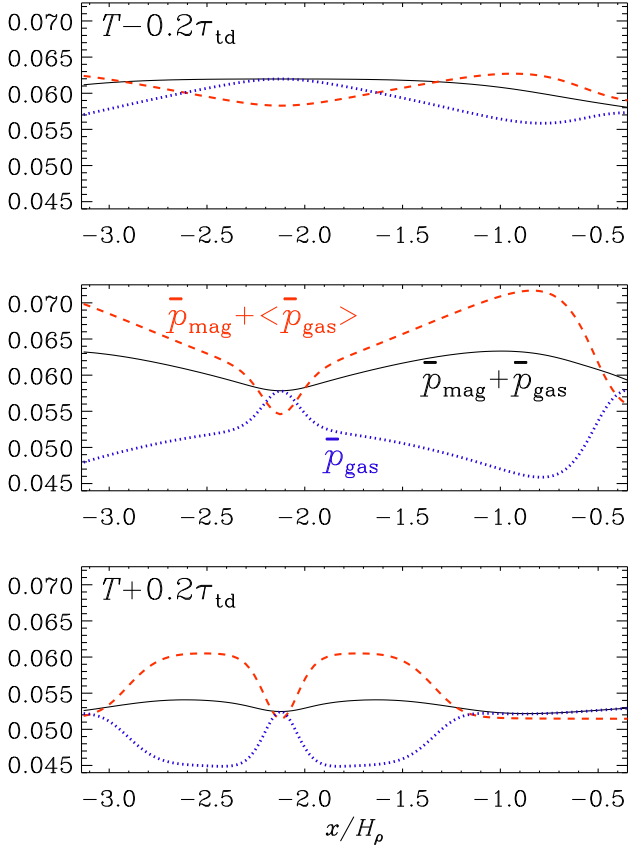}
\end{center}\caption{
2D MFS for case~II' showing the mean pressure balance at the surface
at three times: $t=T-0.2 \tautd$ (upper panel),
$t=T$ (middle panel), and $t=T+0.2 \tautd$ (lower panel),
around the moment when the mean magnetic pressure
gradient is largest (at $t\equiv T$).
}\label{ppgas_pmag}\end{figure}

To demonstrate that the sharp structures are caused by the Lorentz force,
particularly the mean magnetic pressure gradient, we compare in
\Fig{ppgas_pmag} the surface profiles (at $z/H_\rho=\pi$) of mean gas
pressure $\meanp_{\rm gas}=\meanrho\cs^2$, mean magnetic pressure
$\meanp_{\rm mag}=\meanBB^2/2$ (shifted upward by a constant,
which is here the horizontally averaged mean gas pressure at the surface,
$\bra{\meanp_{\rm gas}}$),
and mean total pressure, $\meanp_{\rm gas}+\meanp_{\rm mag}$.
The gas pressure gradient is always directed away from the sharp structure
and opposes its formation.
The mean magnetic pressure gradient is directed toward the sharp
structure and overcomes the mean gas pressure gradient at the time when
the structure forms.

\subsection{Similarity between DNS and MFS}

The sharp structures appear superficially similar to those found
by \cite{MBKR14} and \cite{Jabbari16}.
In the present case, a sharp structure is seen to appear at
$t/\tautd=1.35$ and it disappears already at $t/\tautd=1.45$.
However, in units of $\tautd'$, which are the units used by
\cite{MBKR14} and \cite{Jabbari16}, the corresponding time interval
is of the order of unity and thus compatible with \cite{MBKR14} and
\cite{Jabbari16}, where the sharp structures persist and appear to
``stick'' together for the duration of $\tautd'$.

\begin{figure}\begin{center}
\includegraphics[width=.99\columnwidth]{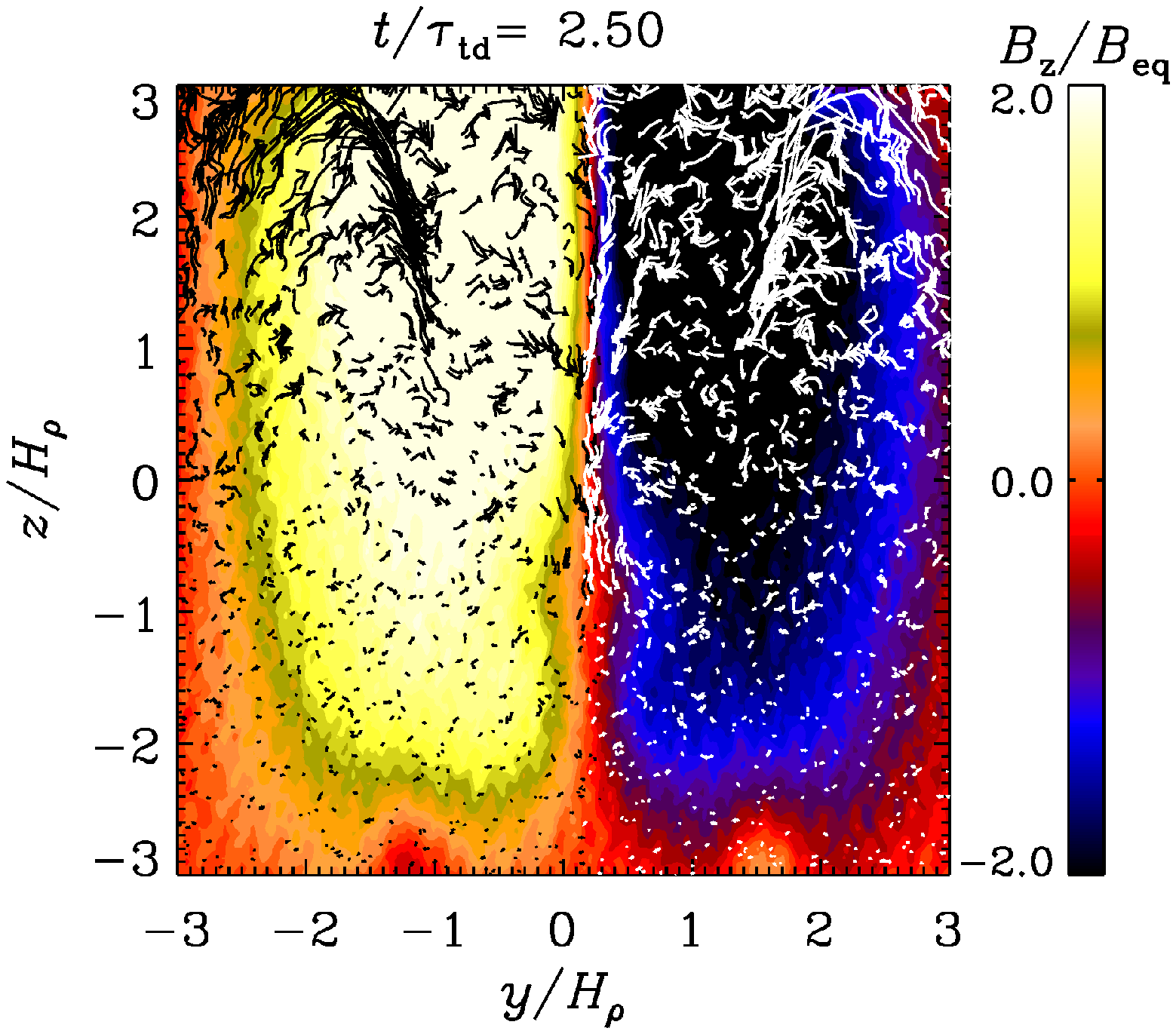}
\includegraphics[width=.8\columnwidth]{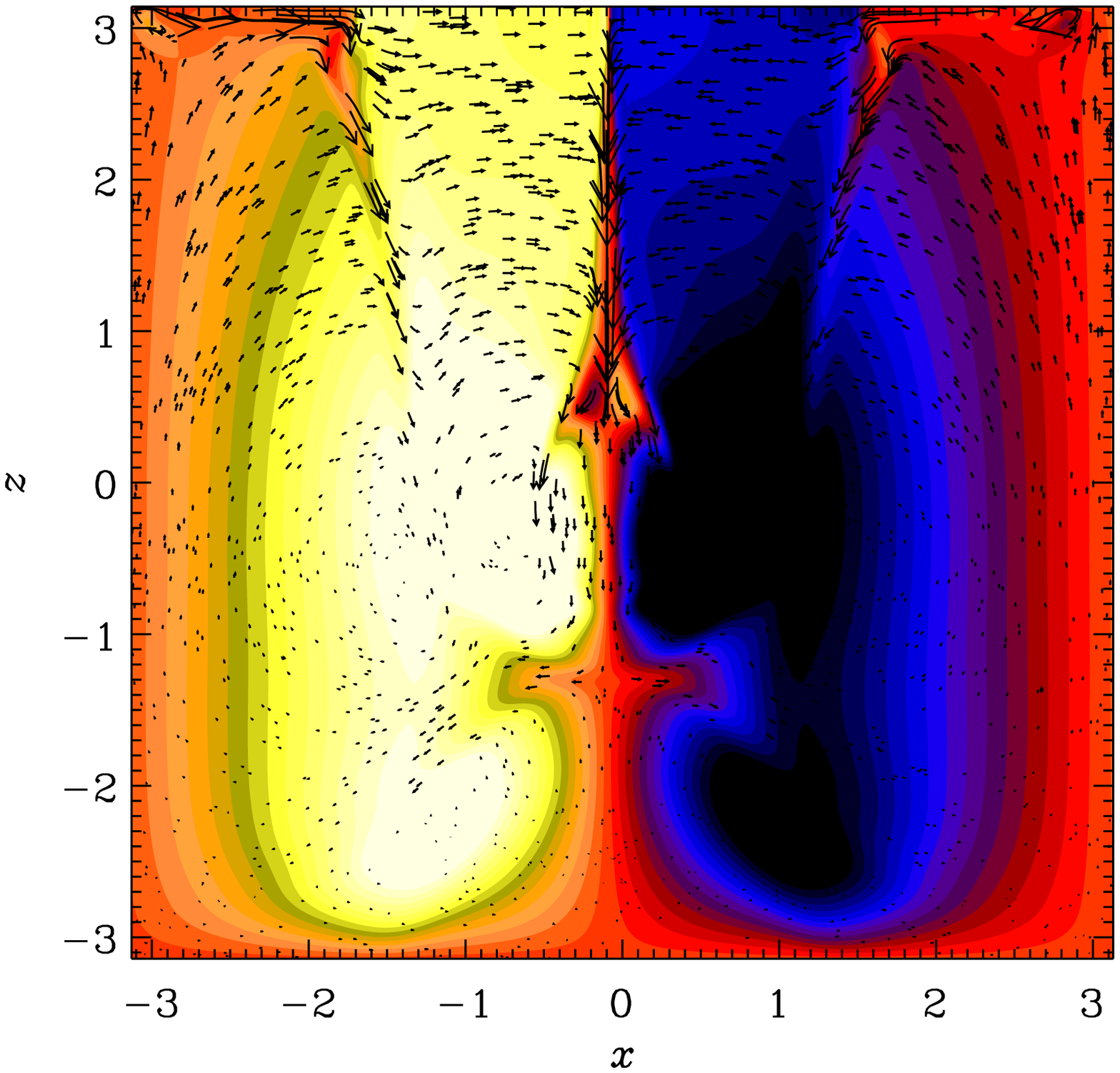}
\end{center}\caption{
$\mean{B_{z}}/\Beq$, together with $\mean{u_{y}}/\Beq$
and $\mean{u_{z}}/\Beq$ vectors for the fully helical run to
compare DNS for $\Rm=50$ (upper panel, Run~R1) with MFS (lower panel, Run~II).
}\label{pvarm_rpi}\end{figure}

We recall that we have assumed $\alpha_0=\const$,
i.e., the $\alpha$ effect is independent of $z$.
This is appropriate for simulating a one-layer system,
in particular Run~RM1zs of \cite{Jabbari16}.
The value of $C_\alpha$ is therefore essentially determined by
the scale separation ratio, $\kf/k_1$; see \cite{Jabbari14}.
Furthermore, the value of $C_u$ can be estimated by using the mean-field
expression $\etaT\approx\urms/3\kf$, which yields $C_u=3\kf/\tilde{k}_1$.
\cite{MBKR14} and \cite{Jabbari16} used $\kf/k_1=30$ and since
$\tilde{k}_1=k_1/4$, their value is $C_u=360$, which is larger than
those considered here.

\FFig{pvarm_rpi} shows a comparison of the magnetic structure together with
velocity vectors in DNS (upper panel) with MFS (lower panel).
One can see that the location of converging flow structures and
downdrafts is similar in both DNS and MFS.

\begin{figure*}\begin{center}
\includegraphics[width=.51\columnwidth]{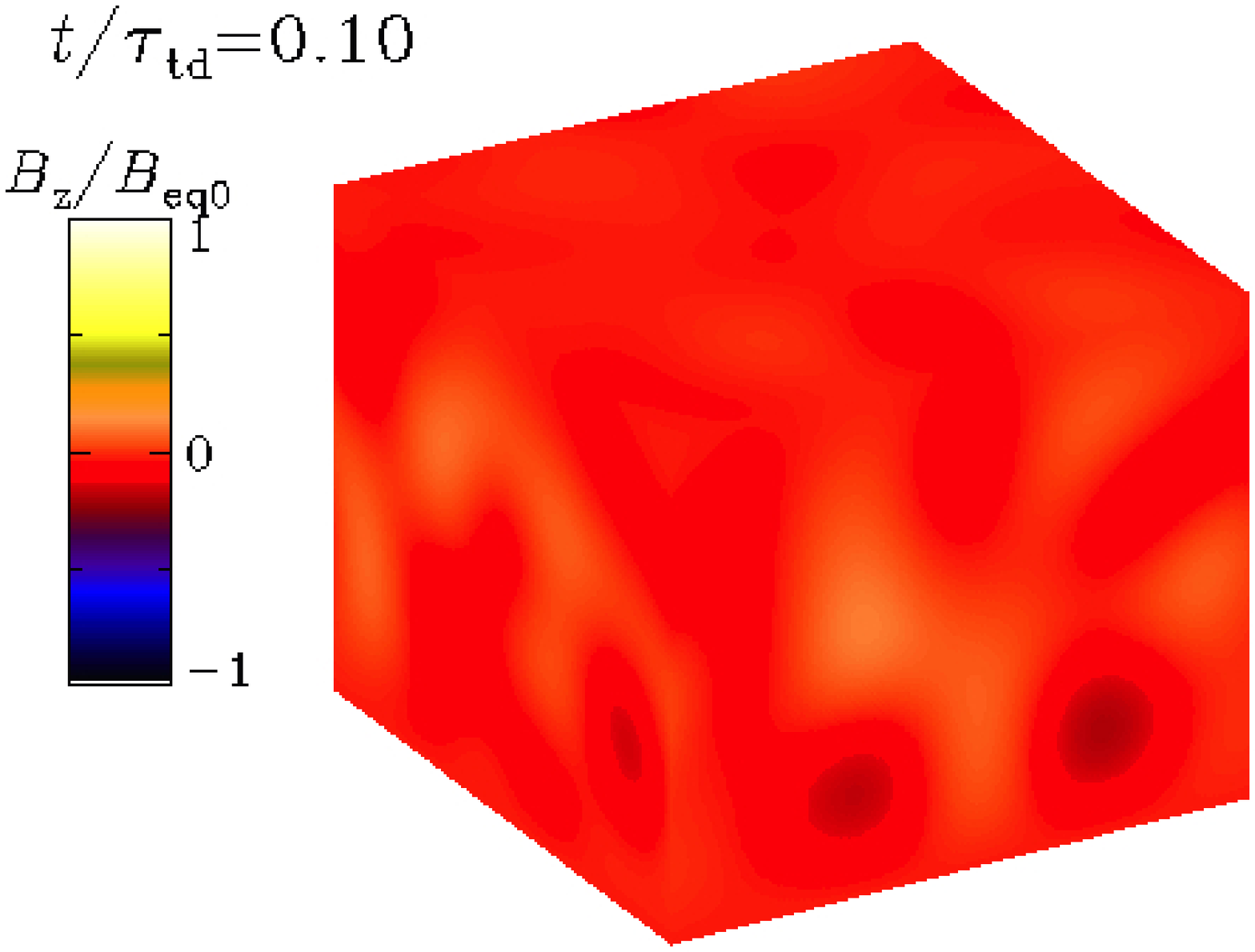}
\includegraphics[width=.51\columnwidth]{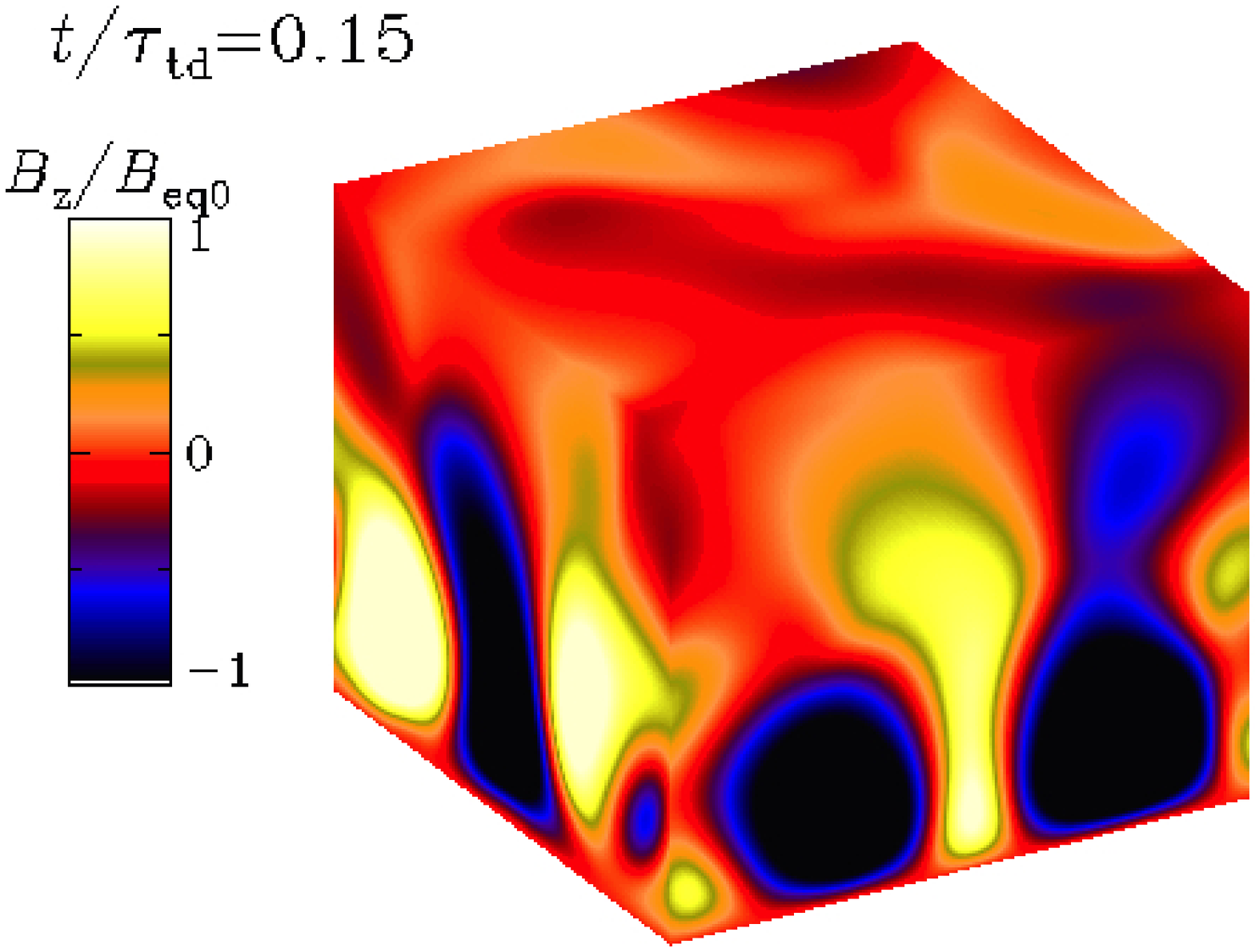}
\includegraphics[width=.51\columnwidth]{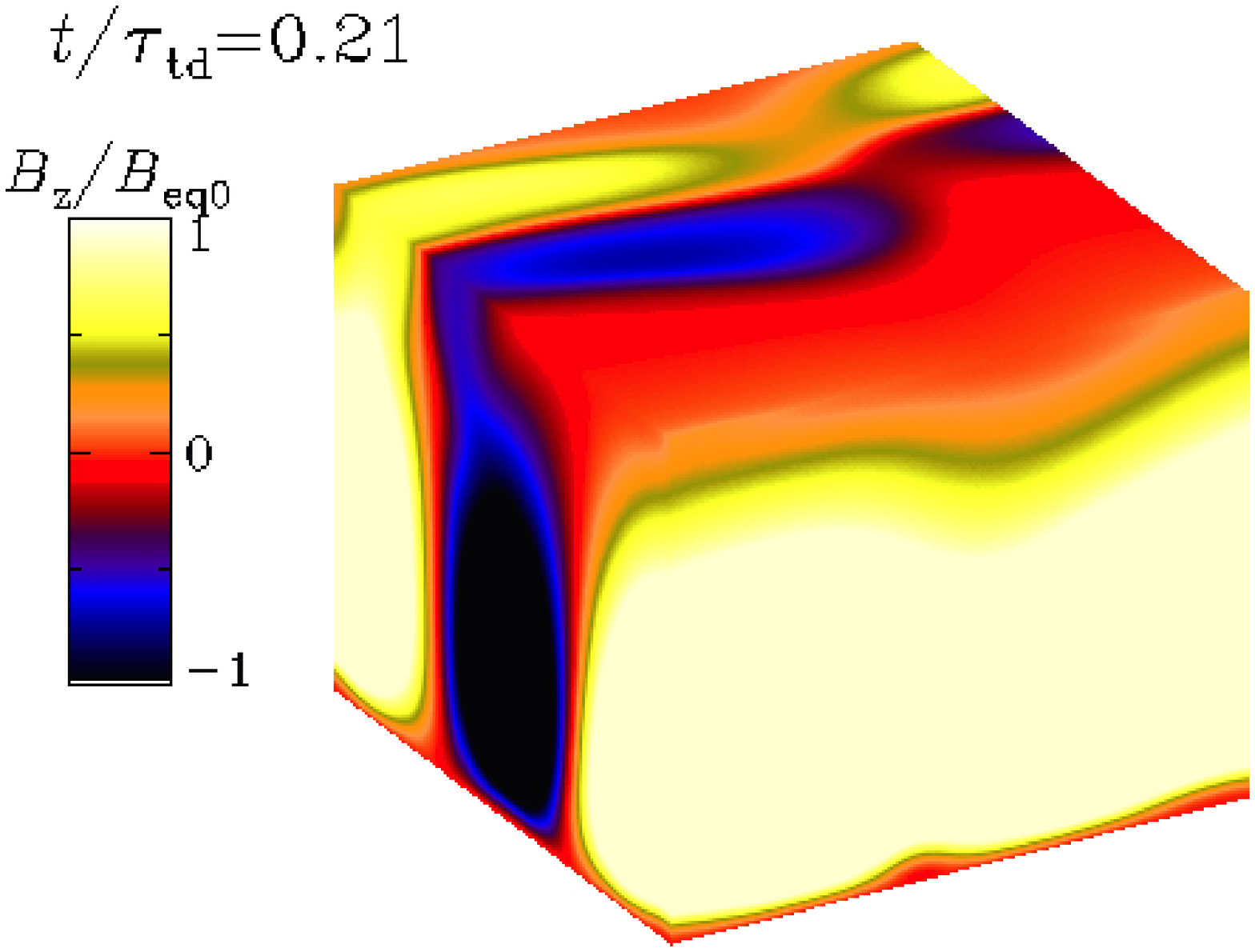}
\includegraphics[width=.51\columnwidth]{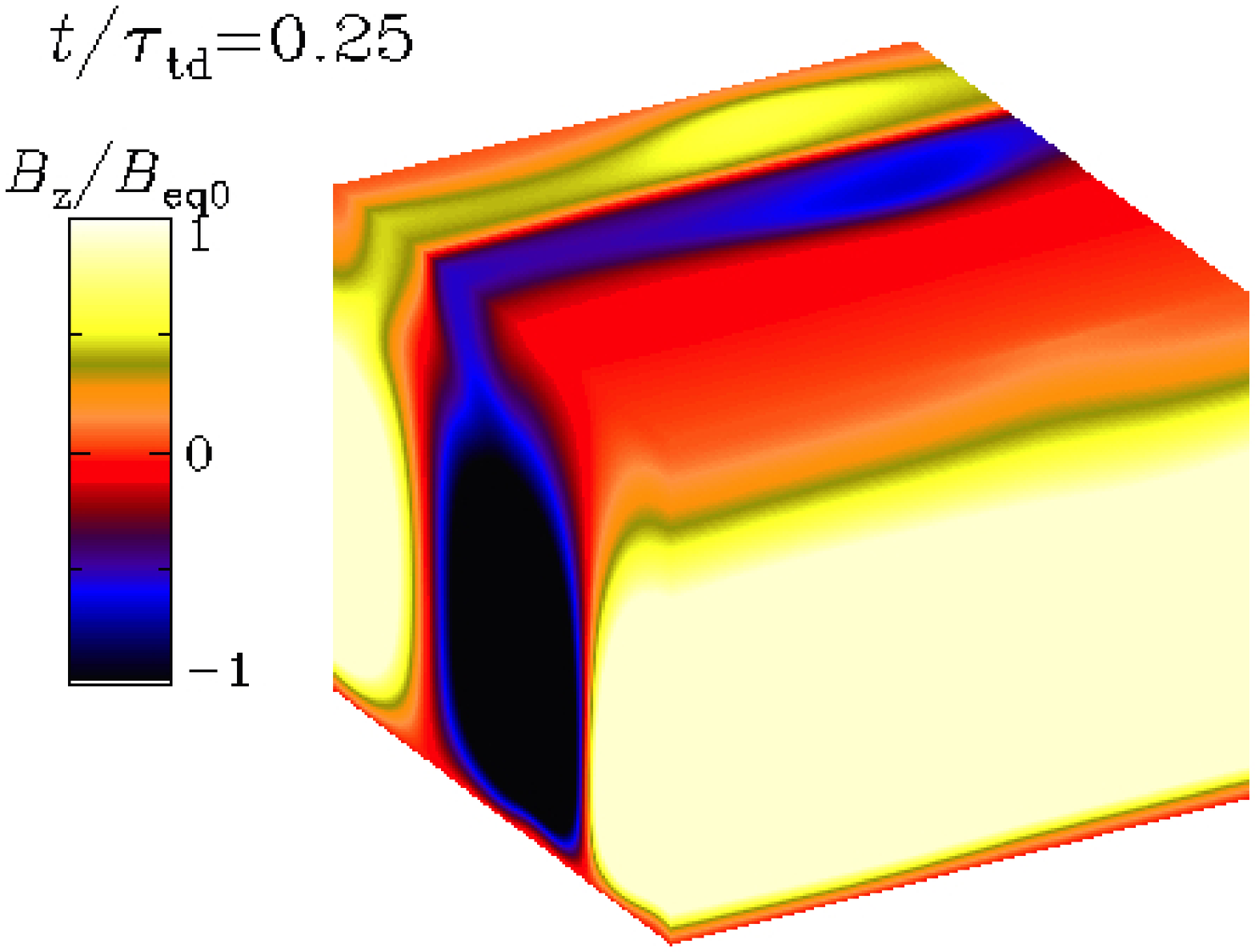}\\
\includegraphics[width=.51\columnwidth]{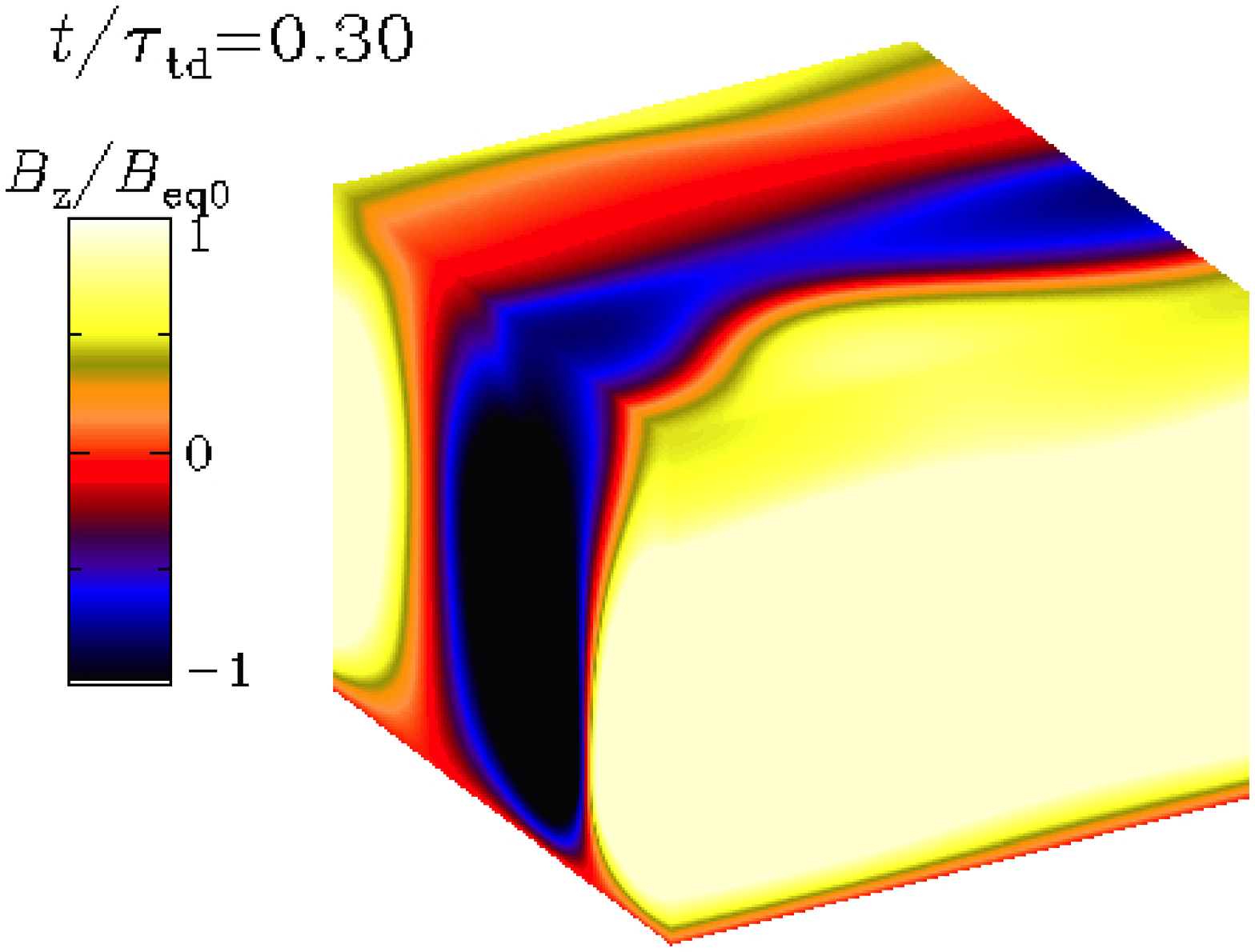}
\includegraphics[width=.51\columnwidth]{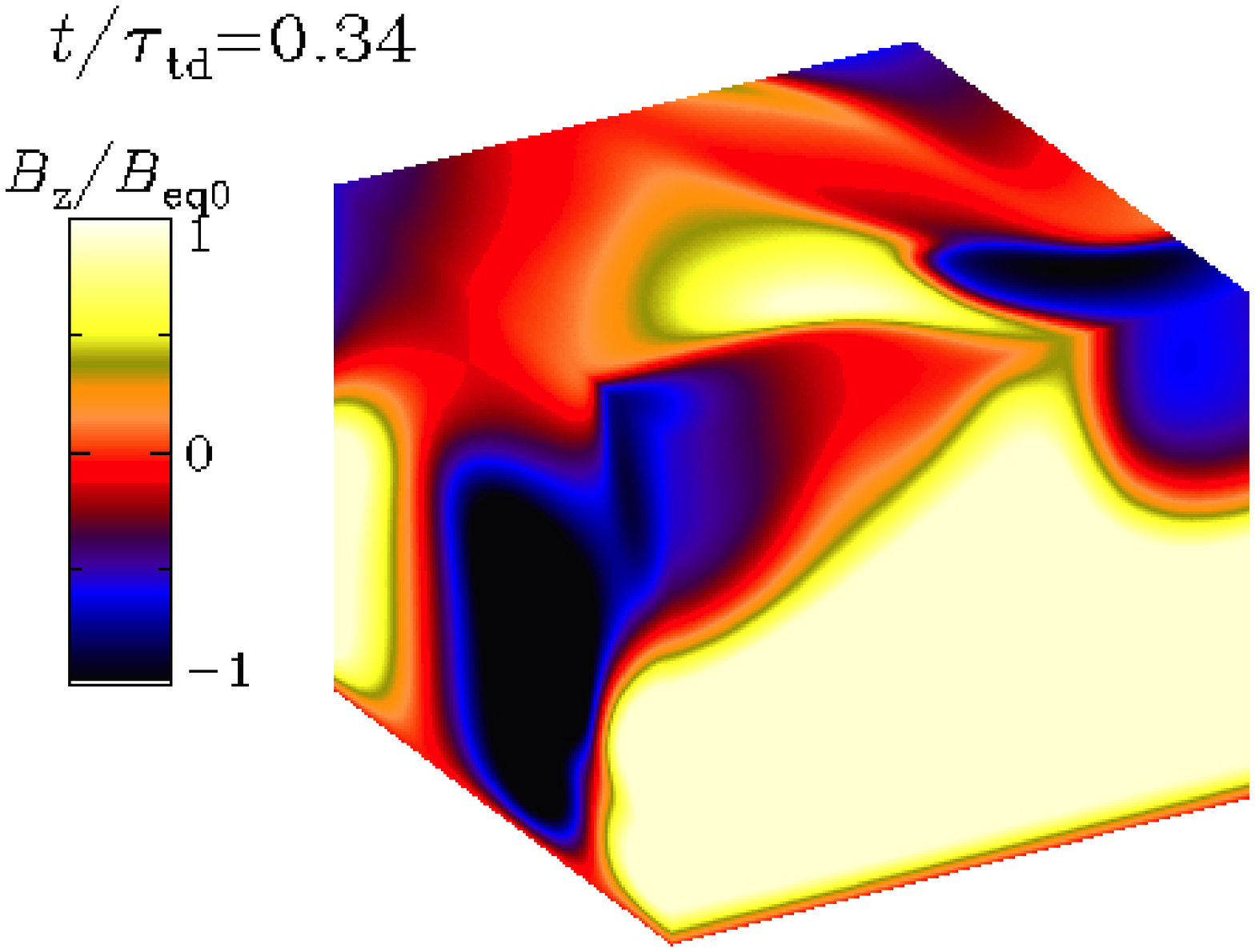}
\includegraphics[width=.51\columnwidth]{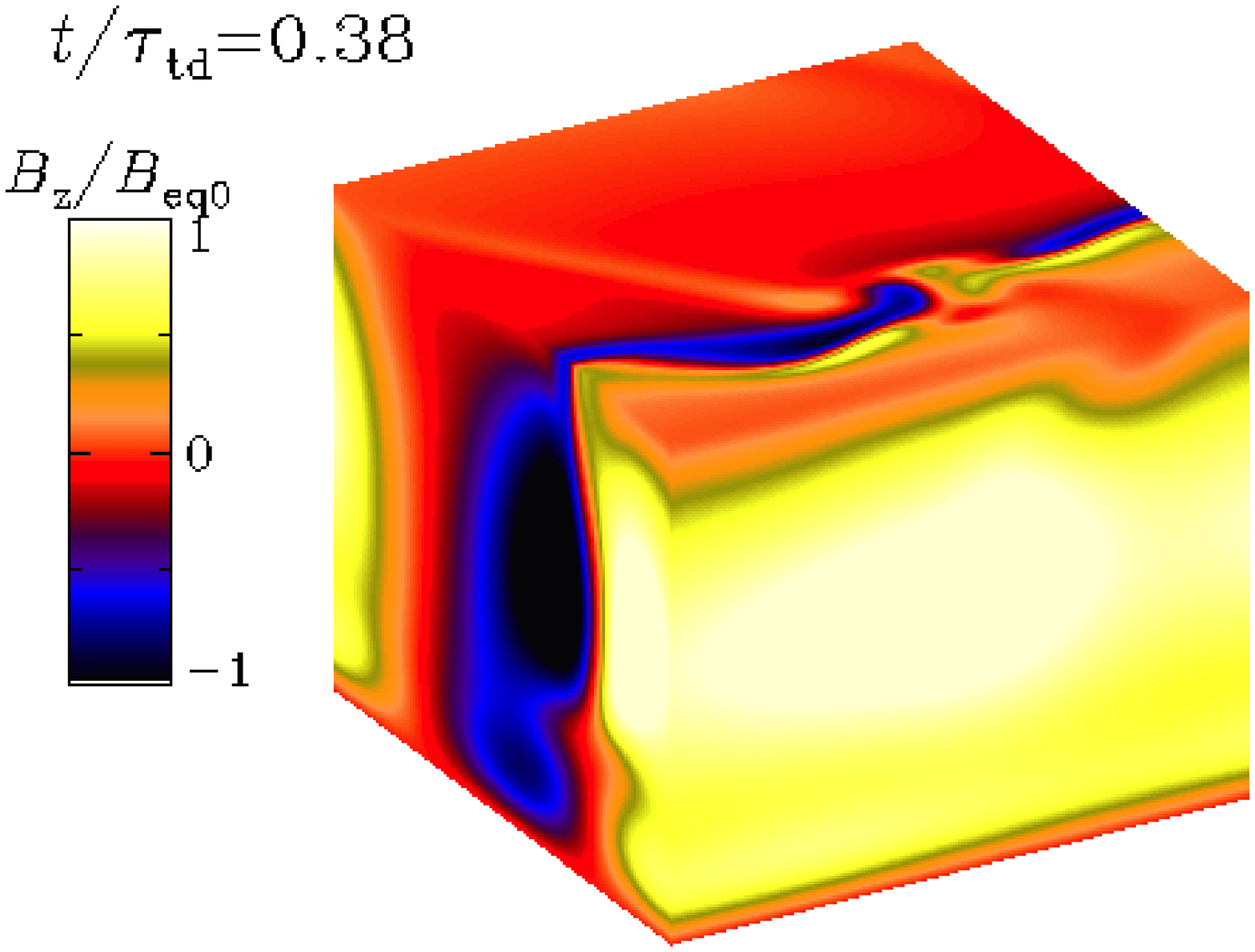}
\includegraphics[width=.51\columnwidth]{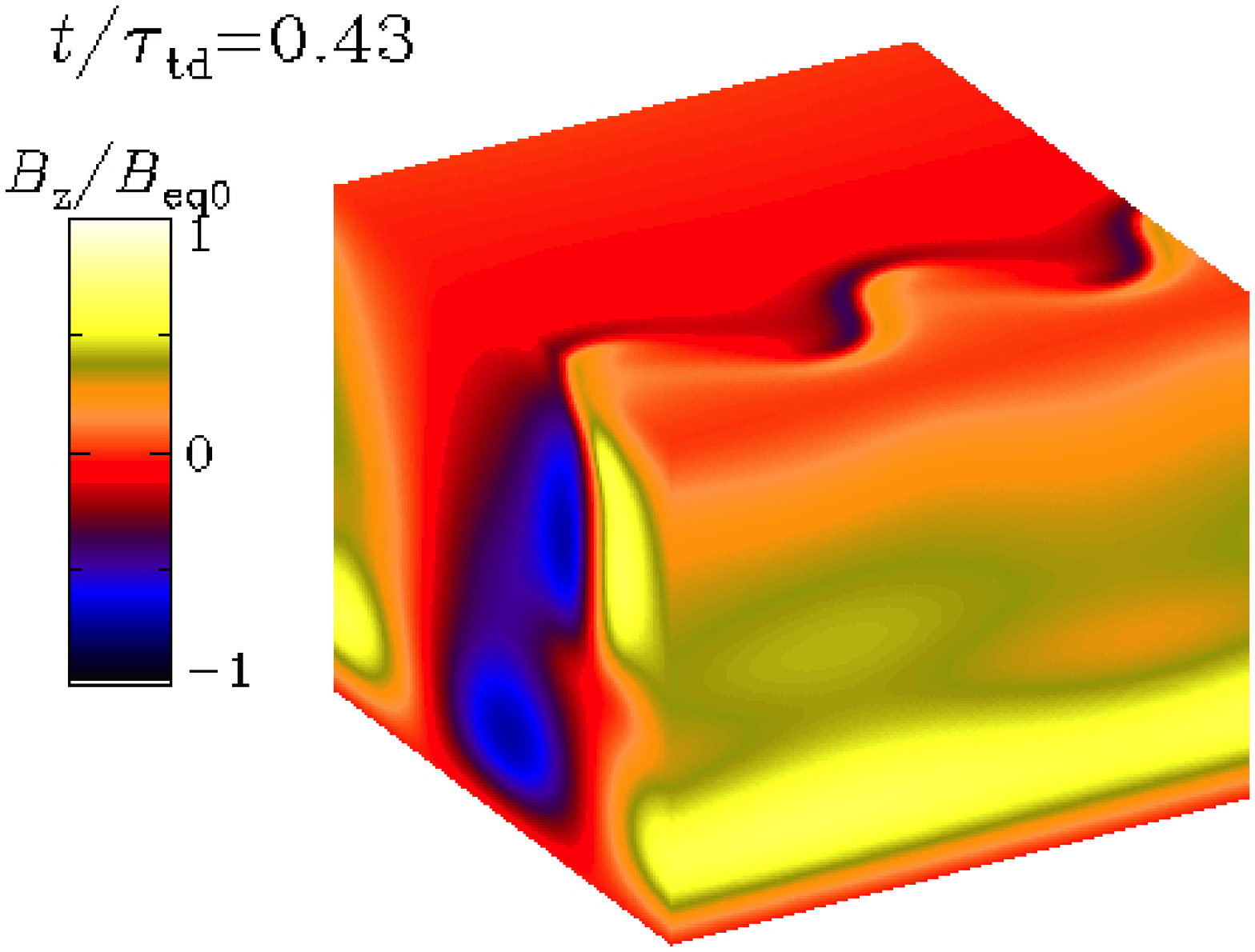}
\end{center}\caption{
Visualization of $\meanB_z/\Beqz$ on the periphery of the computational
domain for Run~I3D (3D MFS).
}\label{box3D}\end{figure*}

\begin{figure}\begin{center}
\includegraphics[width=.9\columnwidth]{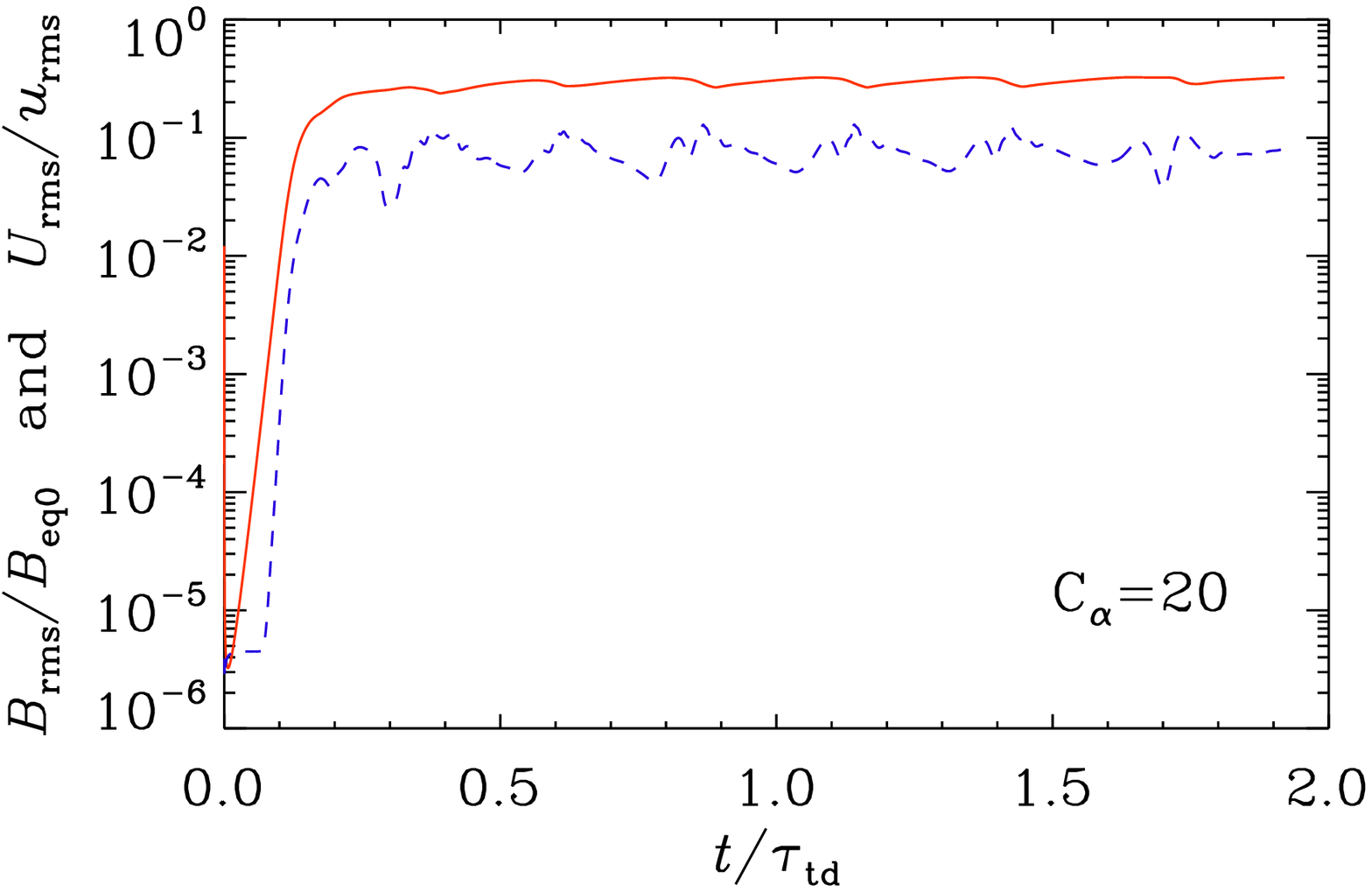}
\end{center}\caption{
$\meanBrms/\Beqz$ (solid red) and $\meanUrms/\urms$ (dashed blue)
vs.\ time for Run~I3D (3D MFS).
}\label{pubrms_3D}\end{figure}

\begin{figure}\begin{center}
\includegraphics[width=.9\columnwidth]{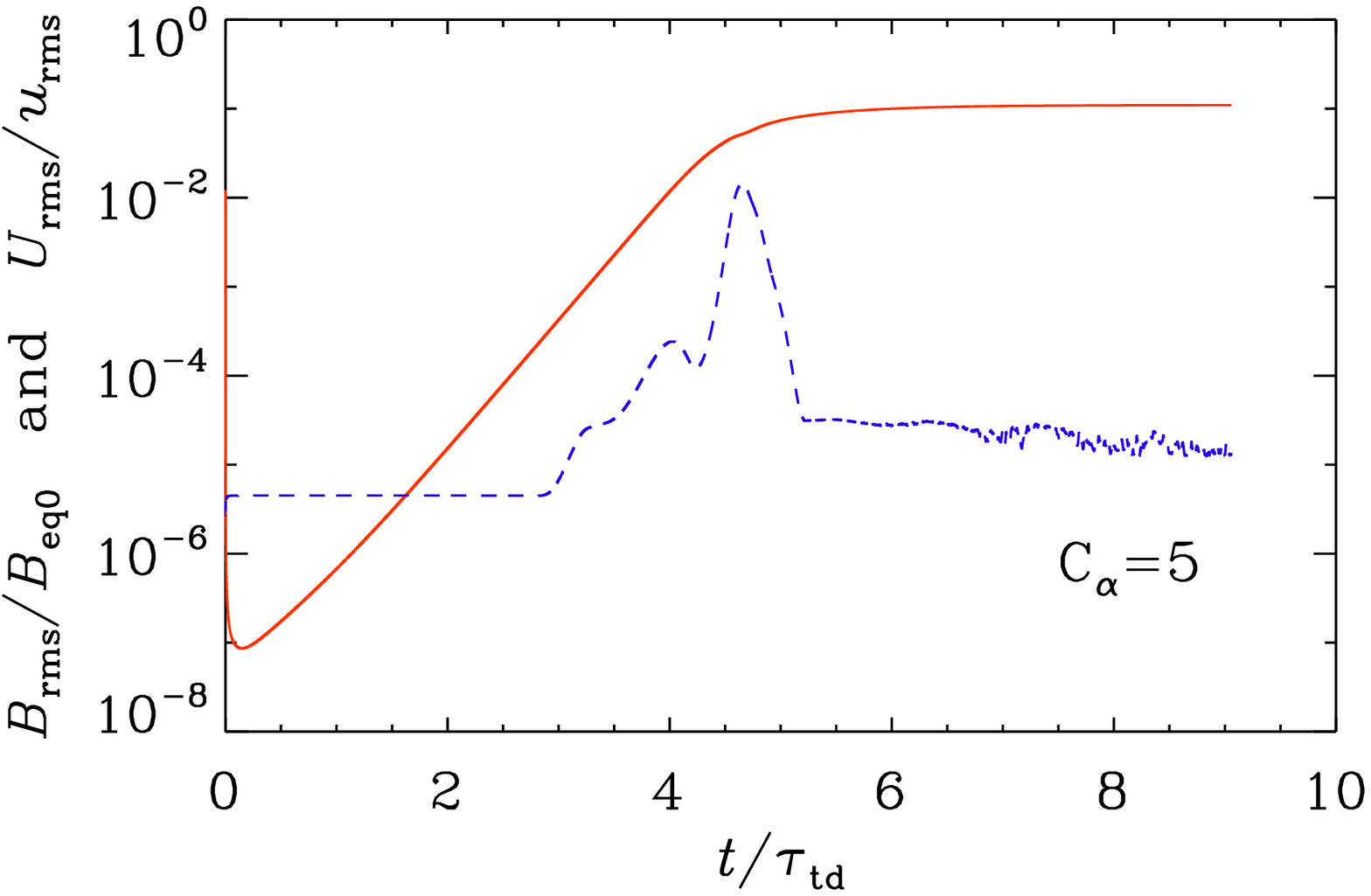}
\end{center}\caption{
$\meanBrms/\Beqz$ (solid red) and $\meanUrms/\urms$ (dashed blue)
vs.\ time for Run~VI3D (3D MFS with weak dynamo).}
\label{pubrms_3D_weak}\end{figure}

\begin{figure}\begin{center}
\includegraphics[width=.9\columnwidth]{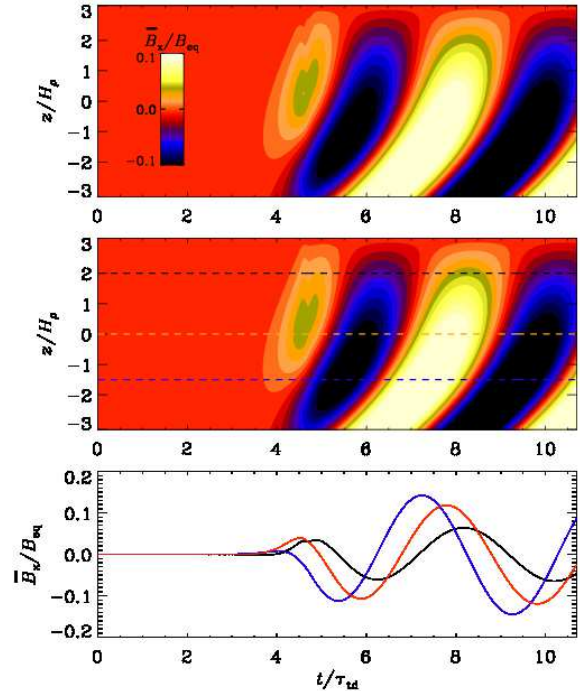}
\end{center}\caption[]{
MFS weak dynamo: butterfly diagrams $\meanB(x_{c},y_{c},z,t)/\Beq(z)$
for Run~VI3D through cross-sections $x_{c}/H_\rho=\pi$, and either
$y_{c}/H_\rho=0.$ (top panel) or $1.8$ (middle panel),
as well as $\meanB_x(x_{c},y_{c},z_{c},t)/\Beq$ through $x_{c}/H_\rho=\pi$
and $y_{c}/H_\rho=1.8$ for $z_{c}/H_\rho=2$ (black line),
$z_{c}/H_\rho=-1.5$ (blue line), and $z_{c}/H_\rho=0.$ (red line)
versus time (lower panel).
In the second panel the dashed black, red, and blue horizontal lines
show the locations where $\meanB_x(x_{c},y_{c},z_{c},t)/\Beq$ is plotted vs.\ $t$.
}\label{Butt_MFS}\end{figure}

\subsection{Three-dimensional MFS}
\label{3DMFS}

As seen in the DNS, the magnetic structures are not two-dimensional
at all times.
It is therefore important to perform mean-fields calculations also
in three dimensions.
The result is shown in \Fig{box3D}, where we plot the $z$ component
of the magnetic field on the periphery of the domain.
Looking at an animation, one can see that magnetic structures rotate
in the counterclockwise direction.
This direction would be the other way around in a model with negative
$\alpha$ effect.
We also see (e.g., at $t/\tautd=0.43$) that the reconnection layers
tend to develop Kelvin-Helmholtz instabilities corresponding to shear flows
that have the same sense as in the two-dimensional simulations.

\begin{figure*}\begin{center}
\includegraphics[width=.41\columnwidth]{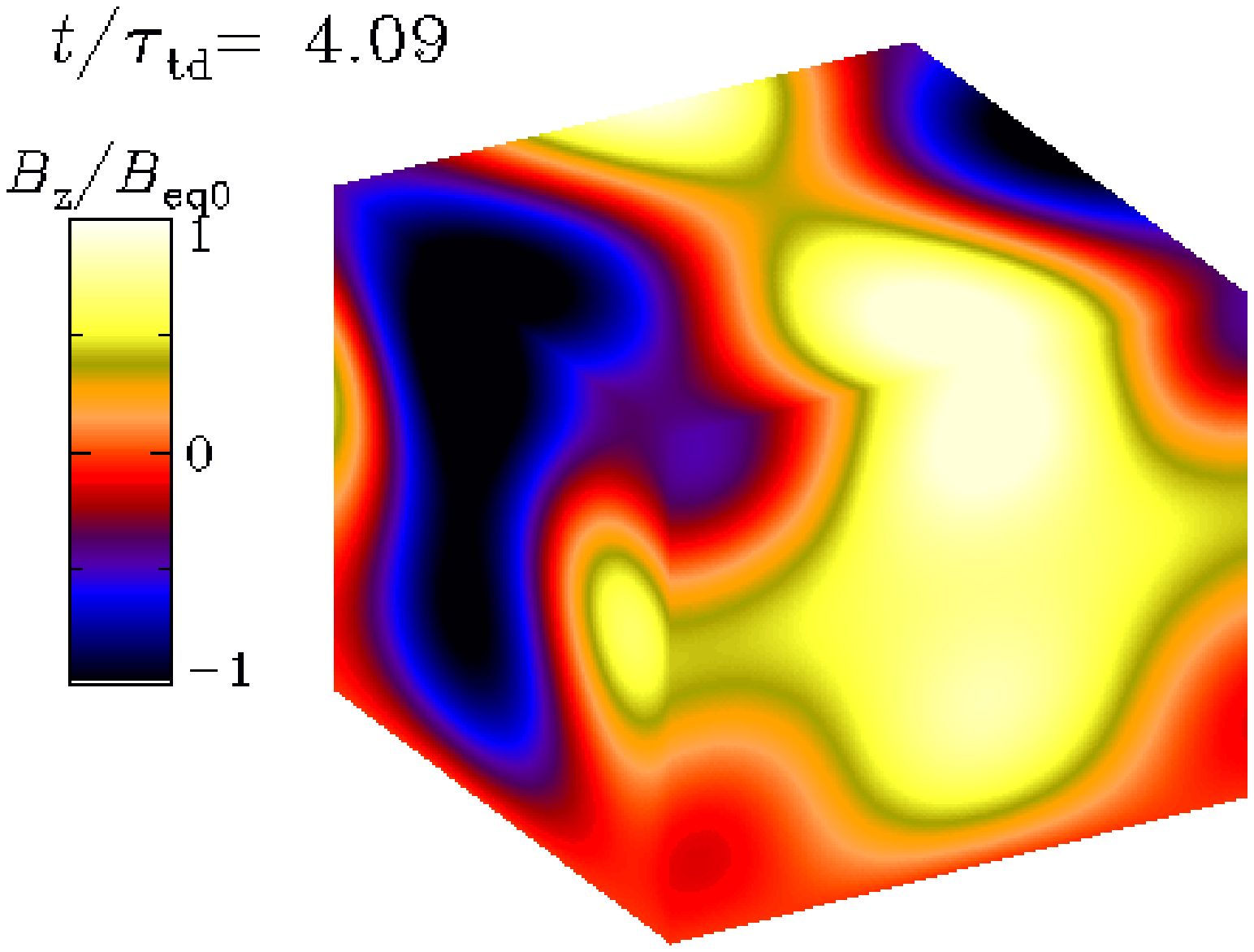}
\includegraphics[width=.41\columnwidth]{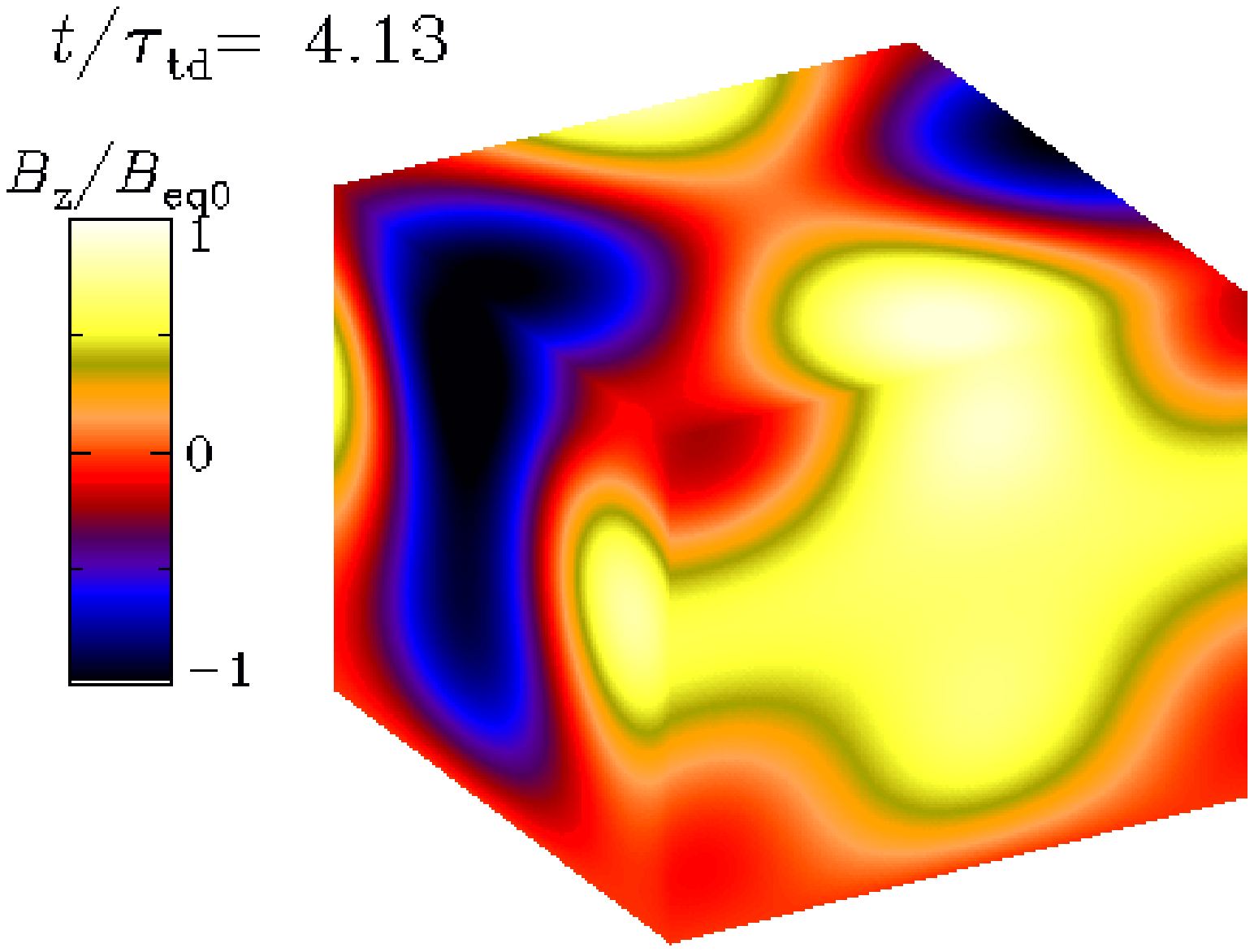}
\includegraphics[width=.41\columnwidth]{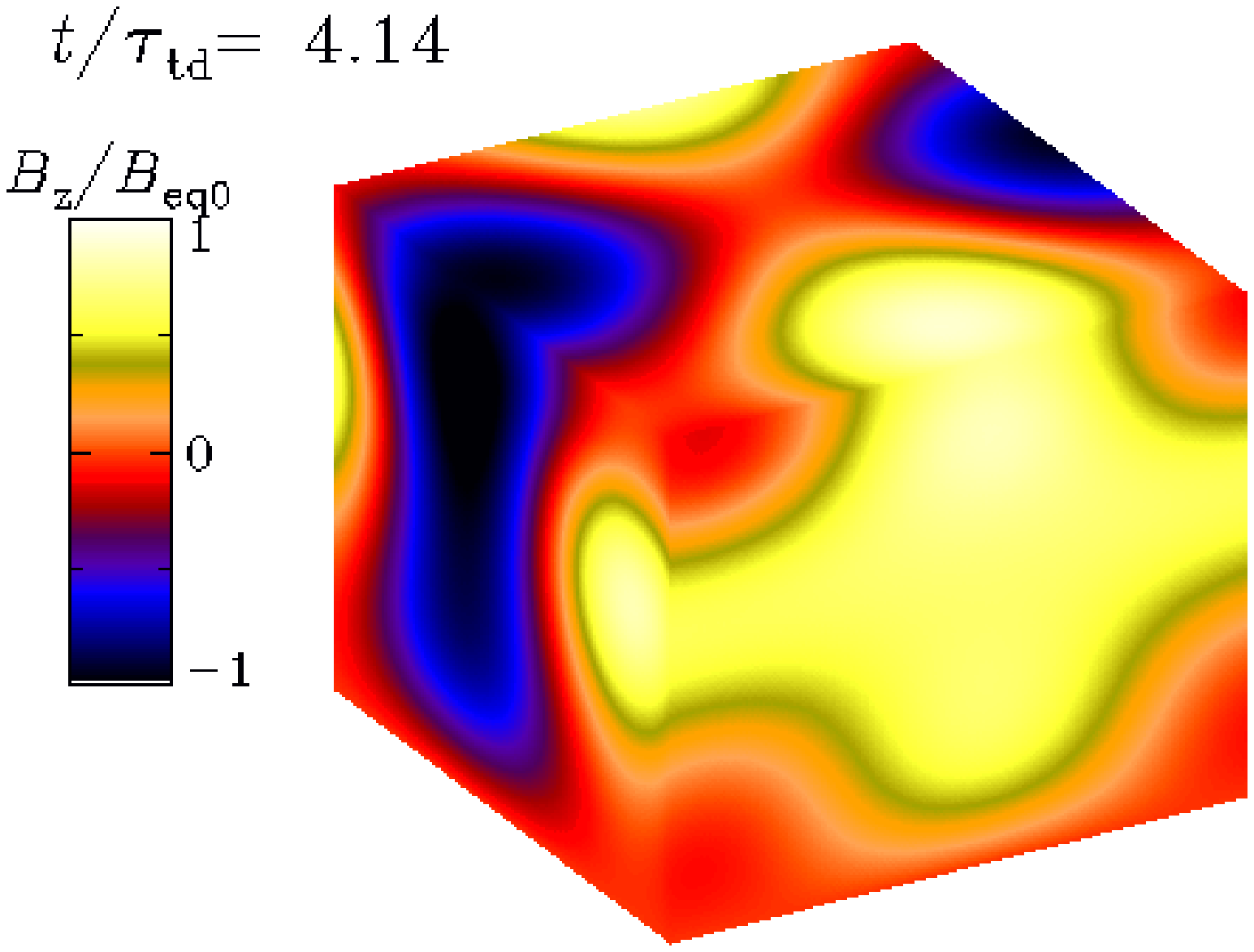}
\includegraphics[width=.41\columnwidth]{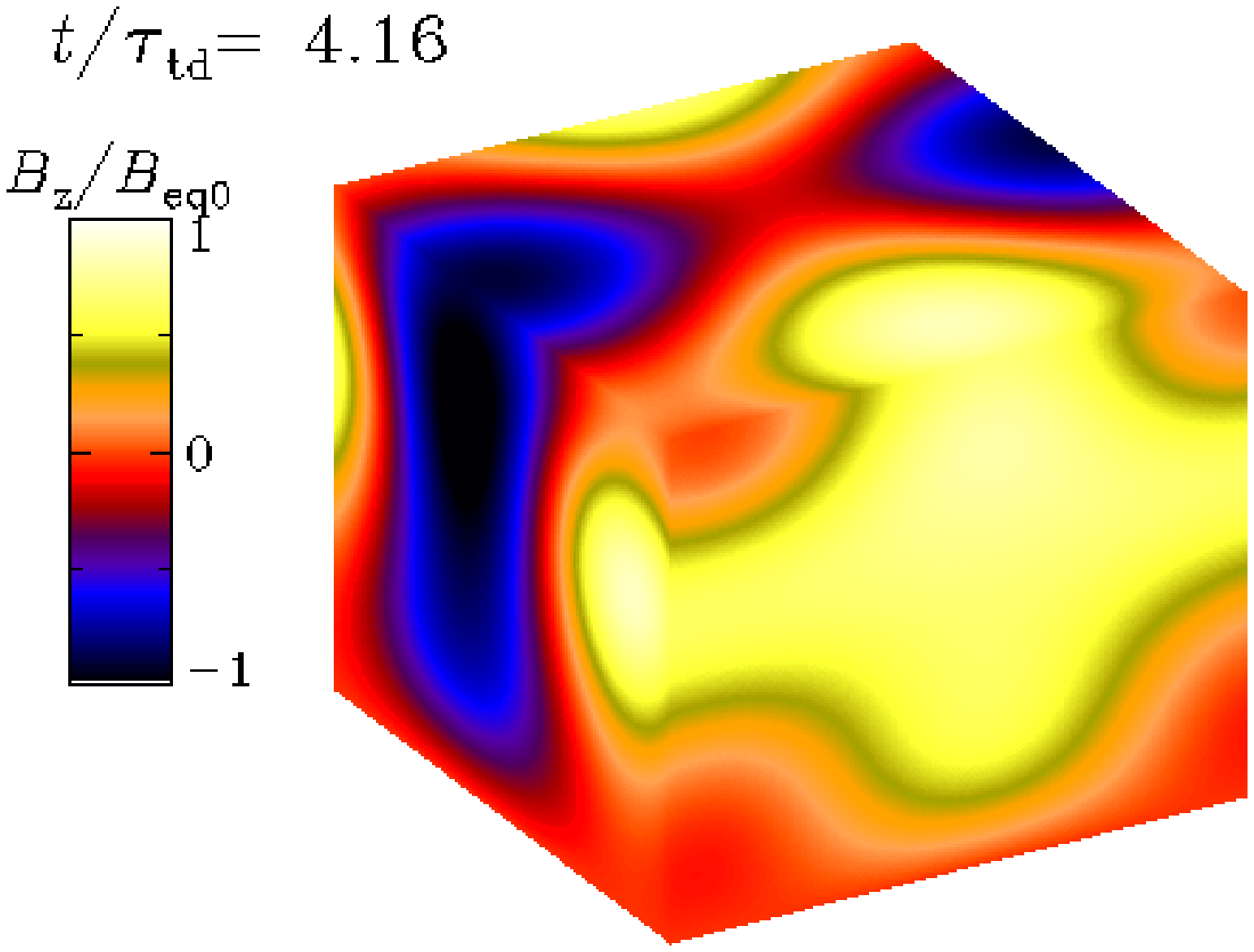}
\includegraphics[width=.41\columnwidth]{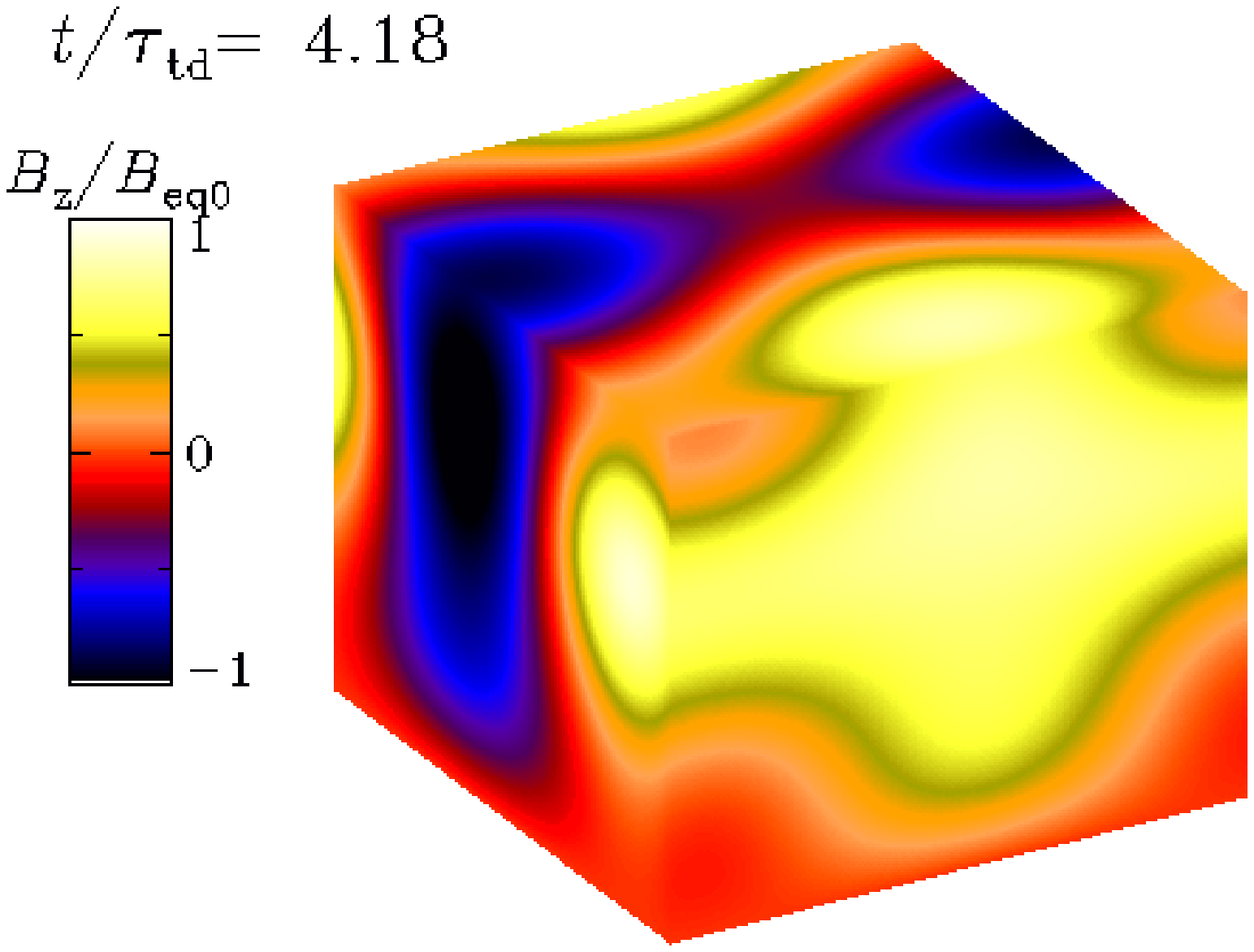}
\end{center}\caption{
Visualization of $\meanB_z/\Beqz$ on the periphery of the computational
domain for Run~VI3D (3D MFS with weak dynamo).}
\label{boxVI}\end{figure*}

\begin{figure*}\begin{center}
\includegraphics[width=.5\columnwidth]{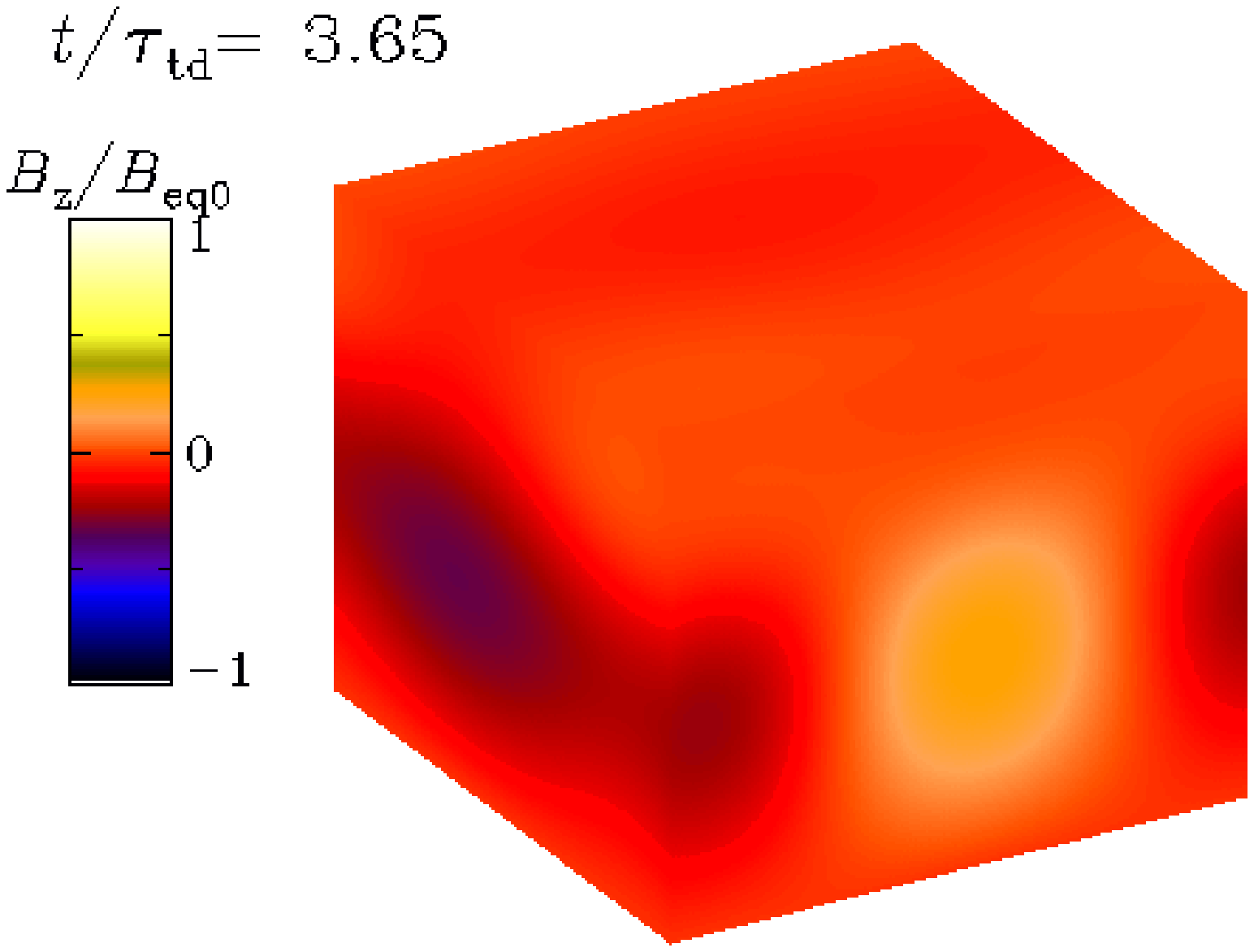}
\includegraphics[width=.5\columnwidth]{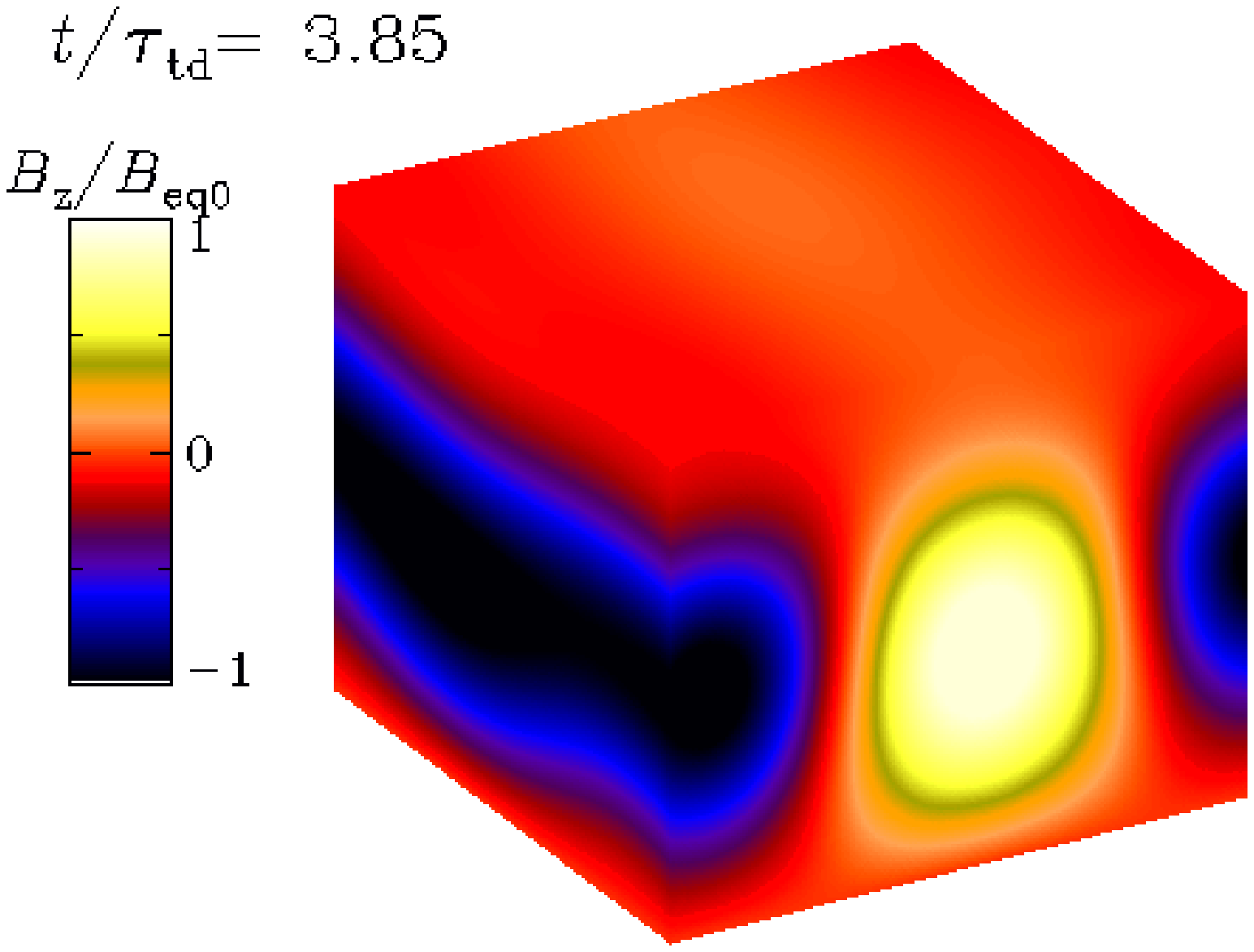}
\includegraphics[width=.5\columnwidth]{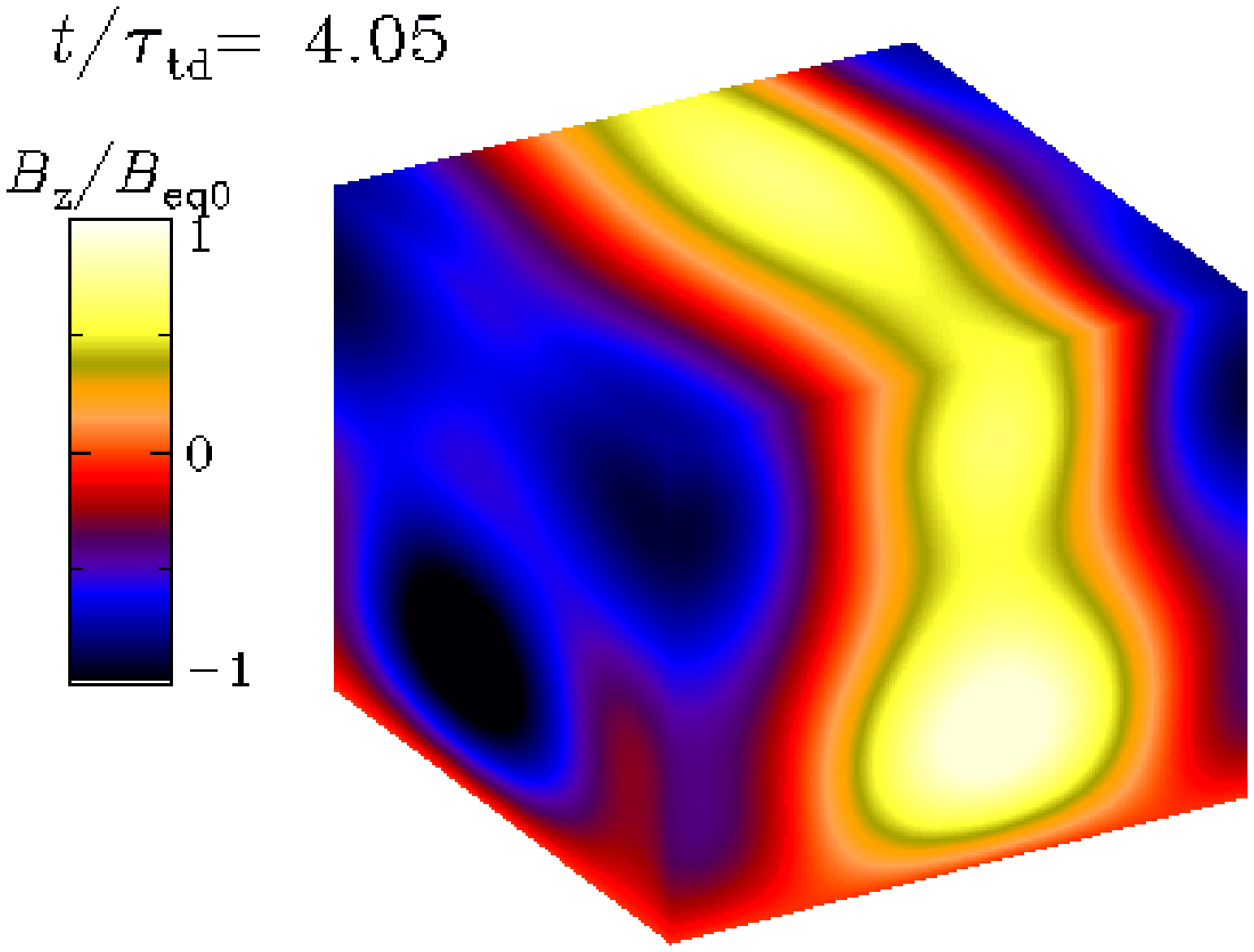}
\includegraphics[width=.5\columnwidth]{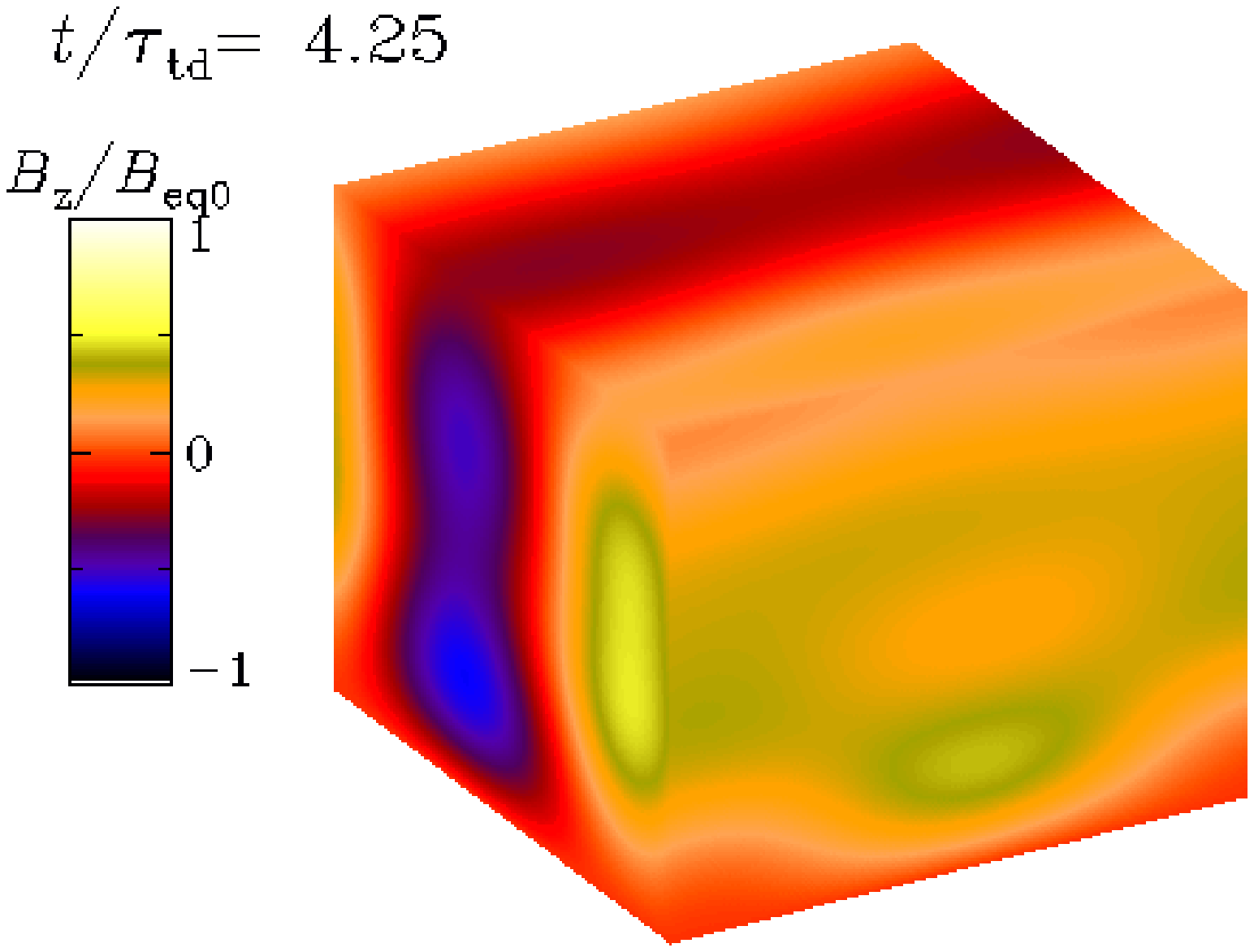}\\
\includegraphics[width=.5\columnwidth]{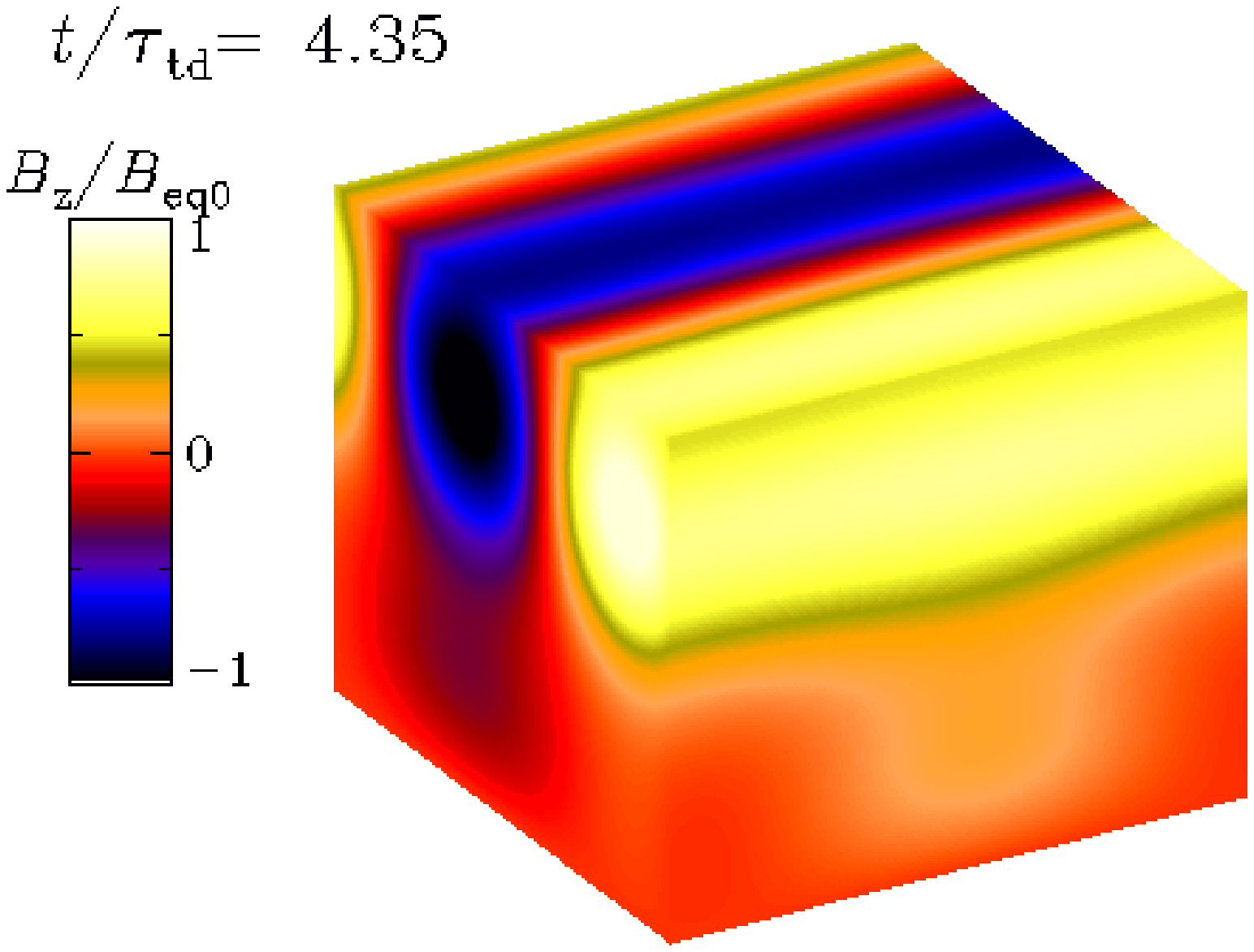}
\includegraphics[width=.5\columnwidth]{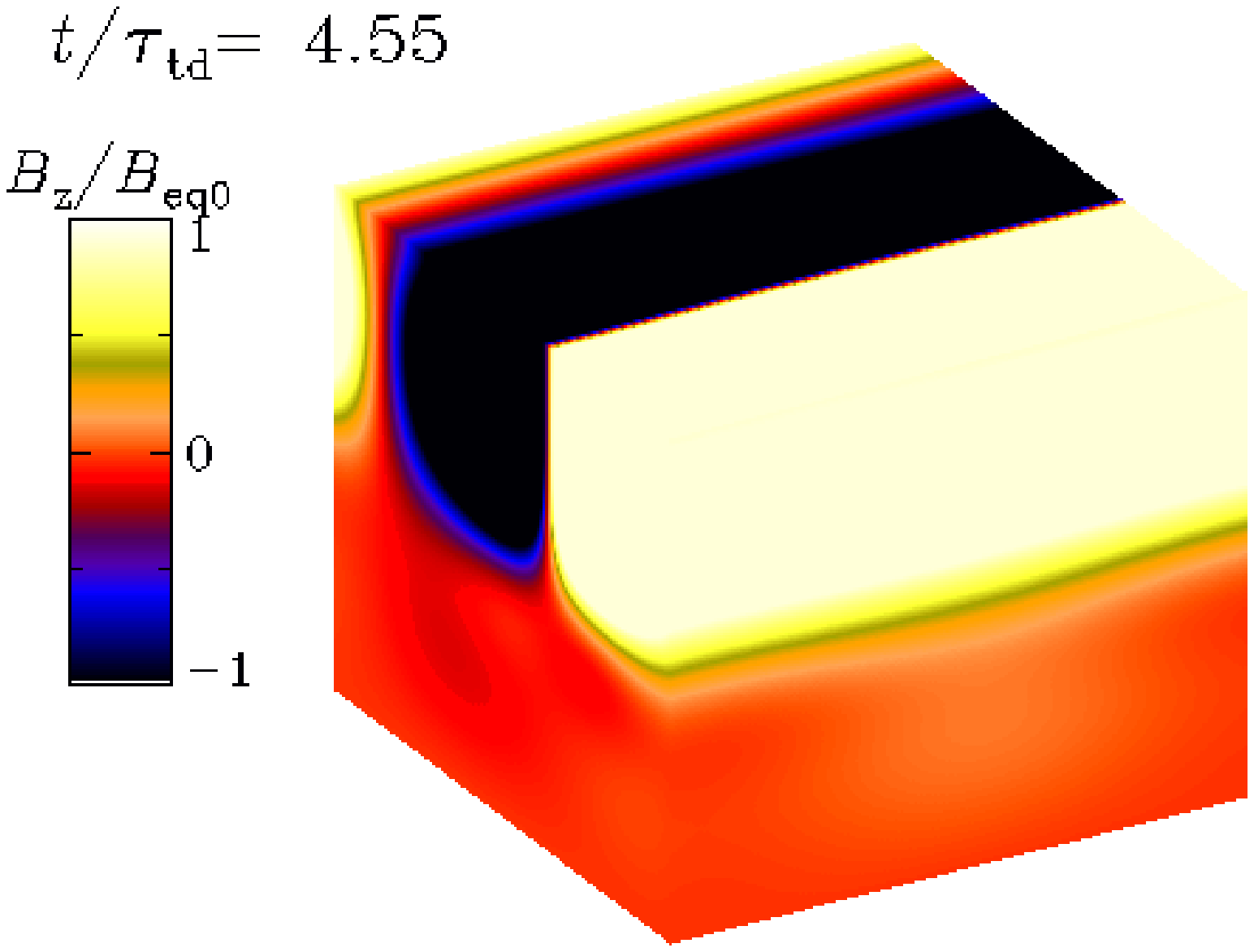}
\includegraphics[width=.5\columnwidth]{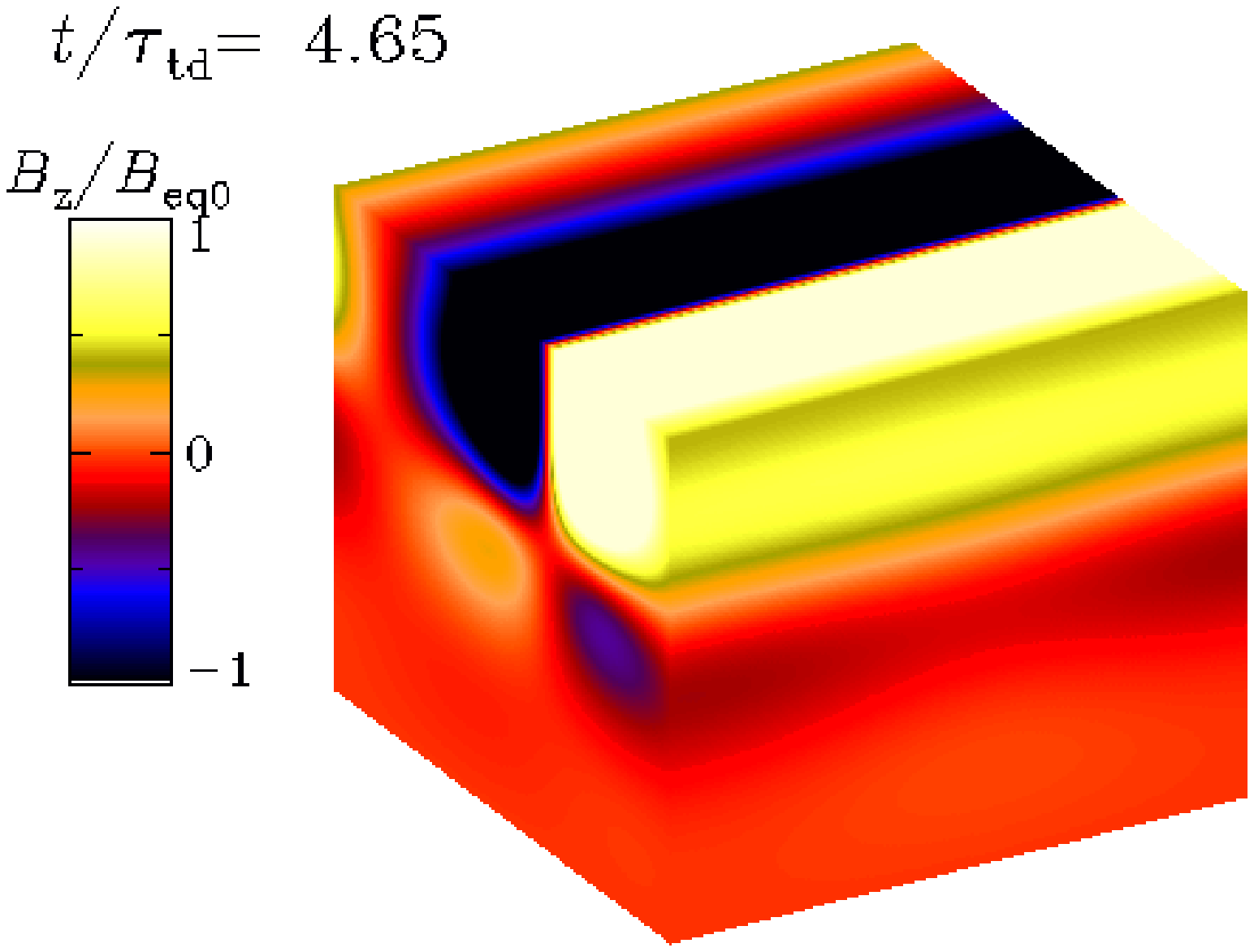}
\includegraphics[width=.5\columnwidth]{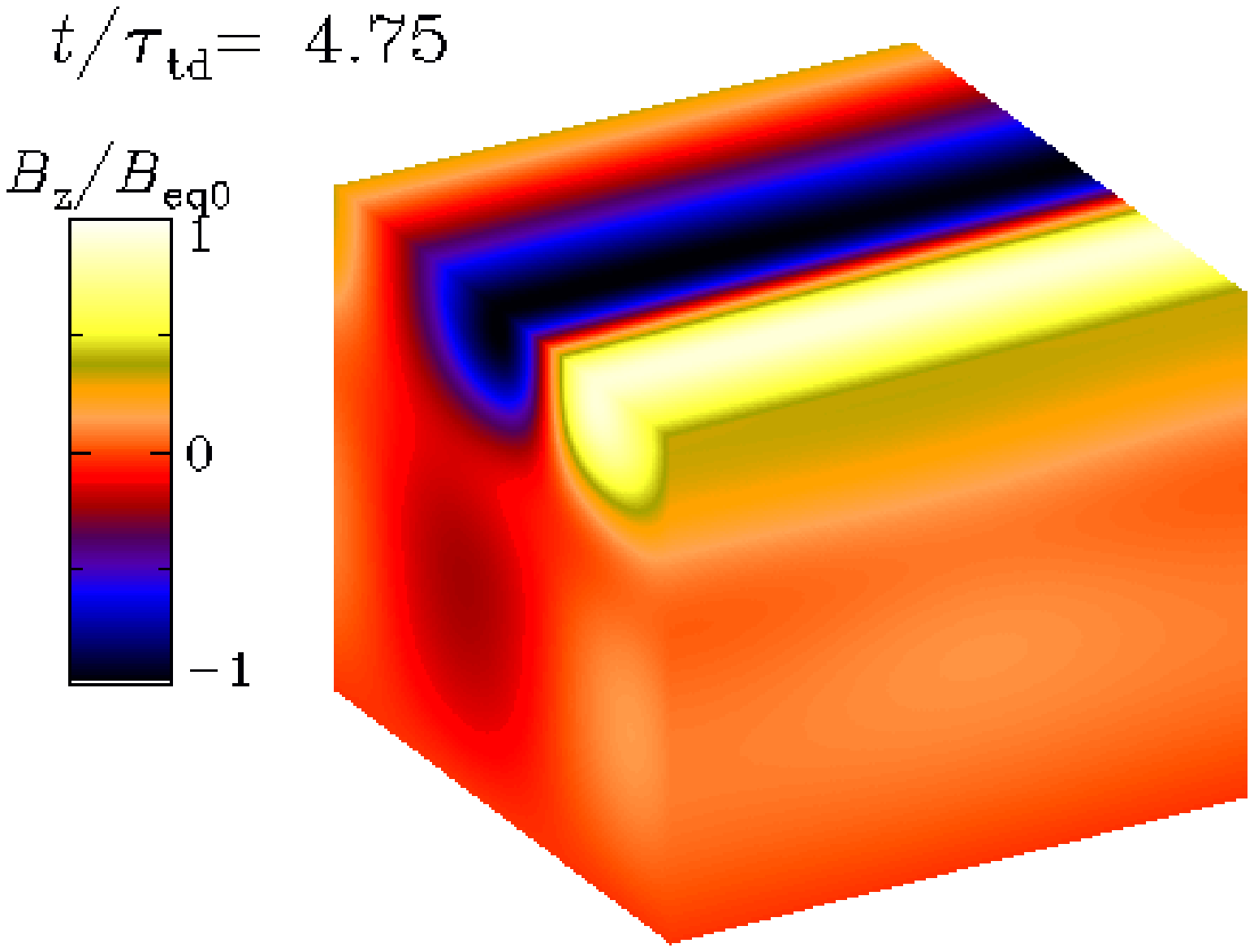}
\end{center}\caption[]{
Time evolution of $\meanB_{z}/\Beq$ in the box for $C_\alpha=5$ with
NEMPI parameterization for Run~VI3D-NEMPI (3D MFS).
}
\label{image_alpha5_nempi}\end{figure*}

Compared with the two-dimensional MFS, the period of oscillations is
now almost three times shorter than in the two-dimensional calculations;
see \Fig{pubrms_3D}.
This is surprising and suggests that the nonlinearity from the feedback
via the mean-field momentum equation is rather important.
Compared with the DNS, the period is larger still.
This nonlinearity might therefore be even more important in the DNS
and that in the MFS the nonlinearity from algebraic $\alpha$ quenching
may be overestimated, i.e., the parameter $Q_\alpha$ was chosen too large.
Compared with \Fig{pubrms}, the minima are now much shallower.

In case VI3D with the lower dynamo number ($C_\alpha=5$, ``weak dynamo''),
the values of $\meanBrms$ and $\meanUrms$ become approximately constant, with
$\meanUrms/\urms$ being much smaller than before; see \Fig{pubrms_3D_weak}.
Nevertheless, the dynamo-generated magnetic field remains oscillatory,
as can be seen from the butterfly diagram in \Fig{Butt_MFS}.
This is consistent with earlier results of \cite{BCC09}.
Another interesting result of employing a lower value of $C_\alpha$
is presented in \Fig{boxVI}, where we show $\meanB_z/\Beqz$
on the periphery of the domain for Run~VI3D.
One can see the clear formation of an {\sf X}-point during the reconnection
of magnetic field lines at the surface of the box (see the third panel).

Next, we perform a 3D MFS study with rotation by including the Coriolis force in
the mean momentum equation (Run~II3D-rot in \Tab{TSum}).
Similarly to the DNS, magnetic structures are formed in the presence of rotation.
As one can see from \Tab{TSum}, the cycle frequency
in this case is two times smaller in comparison with the non-rotating run
with similar parameters.
On the other hand, the cycle frequency observed in the 3D rotating MFS
is nearly the same as in the 2D MFS with similar parameters.

\subsection{3D MFS with NEMPI}
\label{MFS-NEMPI}
To compare with the 3D MFS described in \Sec{3DMFS},
we now include the parameterization of NEMPI
by the following replacement of the mean Lorentz force
in \Eq{MNSE}.
\EQ
\meanJJ\times\meanBB\to\meanJJ\times\meanBB+\nab\left(\half\qp\meanBB^2\right),
\EN
where $\qp$ determines the turbulence contribution to the large-scale
Lorentz force.
Here, $q_{\rm p}$ depends on the local field strength
and is approximated by \citep{KBKR12}
\EQ
\qp(\beta)={\qpz\over1+\beta^2/\betap^2},
\label{qp-apr}
\EN
where $\qpz$ and $\betap$ are constants, and
$\beta=|\meanBB|/\Beq$ is the normalized mean
magnetic field.
For $\Rm\la60$, \cite{BKKR12} found $\qpz\approx 32$
and $\betap\approx 0.058$.

NEMPI describes the formation of magnetic structures through a strong
reduction of turbulent pressure by the large-scale magnetic field.
For large magnetic Reynolds numbers, this suppression of the turbulent
pressure can be strong enough so that the effective large-scale magnetic
pressure (the sum of non-turbulent and turbulent
contributions to the large-scale magnetic pressure) can become negative.
This results in the excitation of a large-scale hydromagnetic instability, namely NEMPI.
In Fig.~\ref{image_alpha5_nempi} we show the
time evolution of $\meanB_{z}/\Beq$ on the periphery of the computational domain
for $C_\alpha=5$ with the NEMPI parameterization included (Run~VI3D-NEMPI).
In this case we also observe the formation of bipolar magnetic structures,
but now mainly in the upper parts of the computational domain.

\subsection{Combined effect of rotation and stratification in DNS}

We now investigate a system of stratified,
non-helically forced turbulence in the presence of rotation.
We study the generation of a large-scale magnetic field driven by
the $\alpha$ effect as a result of the combined effects of rotation
and stratification.
\cite{Jabbari14} have studied a similar system, but in their case
there was also a weak imposed horizontal magnetic field.
Table~\ref{Tab3} shows all non-helical DNS runs with their parameters.

\begin{table}\caption{
Summary of the DNS with rotation and non-helical forcing.
The reference run is shown in bold.
}\vspace{12pt}\centerline{\begin{tabular}{lrlcccccl}
Run & $\Omega$&$\Co$&$\theta$&$\Rm$& $\tilde{\lambda}$\\
\hline
Rn1   &   2  &   1.4 &    0   &  58  &  0.097 \\ 
Rn2   &   2  &   1.4 &   180  &  56  &  0.098 \\ 
Rn3   &   2  &   1.4 &   45   &  61  &  0.087 \\ 
Rn4   &   2  &   1.4 &   90   &  65  &  0.086 \\ 
Rn5   &   4  &   2.8 &    0   &  68  &  0.113 \\ 
{\bf Rn6} & {\bf 8} & {\bf 5.6} & {\bf 0} & {\bf 81} & {\bf 0.107} \\ 
Rn7   &  10  &   6.7 &    0   &  99  &  0.098  \\ 
\label{Tab3}\end{tabular}}
\end{table}

\begin{figure}\begin{center}
\includegraphics[width=\columnwidth]{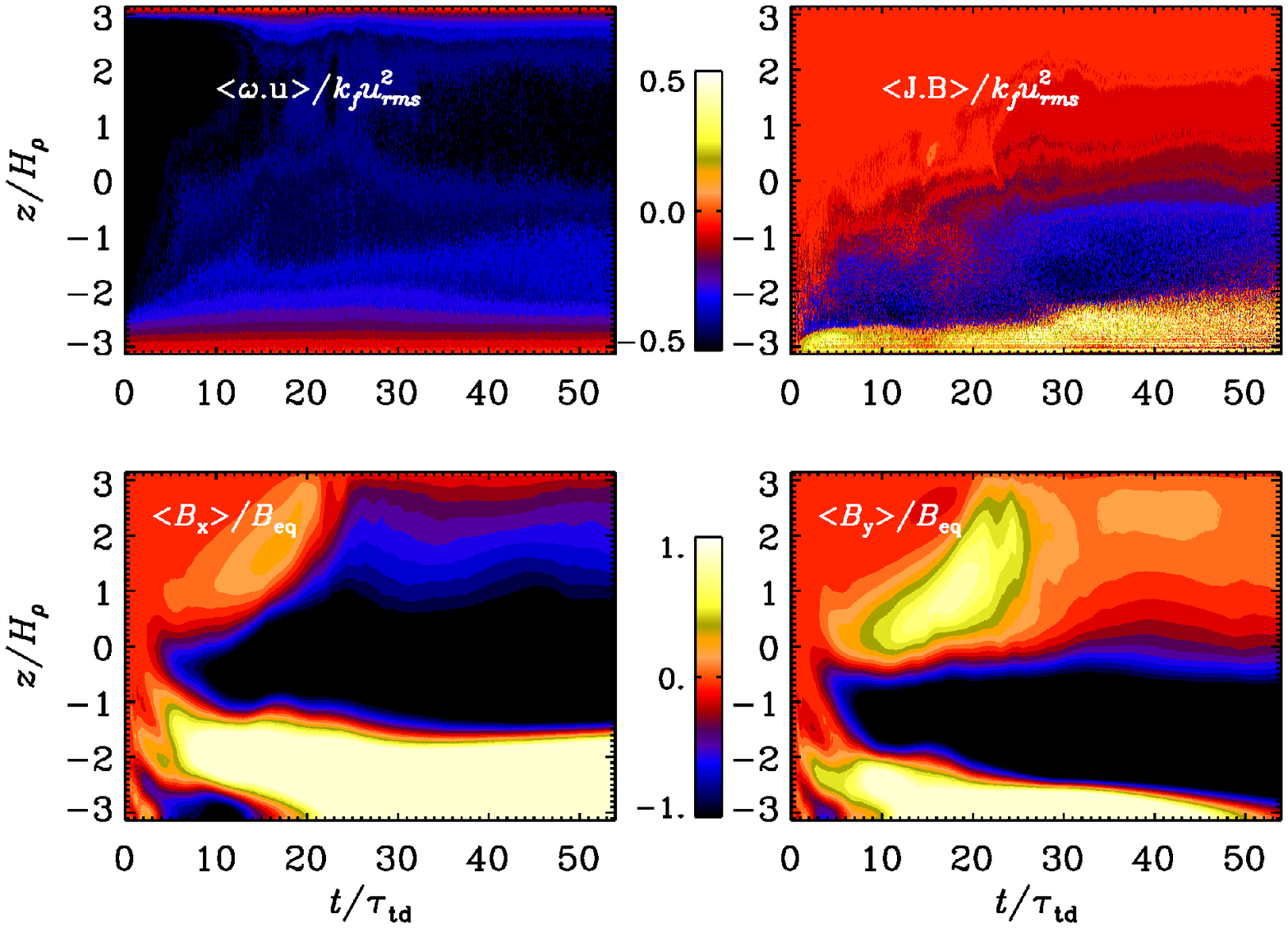}
\end{center}\caption{
$\bra{\oo\cdot\uu}/\kf\urms^2$, $\mu_0\bra{\JJ\cdot\BB}/\kf\urms^2$,
$\bra{B_x}/\Beq(z)$, and $\bra{B_y}/\Beq(z)$ as a function of $z/H_\rho$ and
$t/\tautd$ for Run~Rn6 (DNS of non-helically forced rotating turbulence).}
\label{pxyav56}\end{figure}

\begin{figure}\begin{center}
\includegraphics[width=.8\columnwidth]{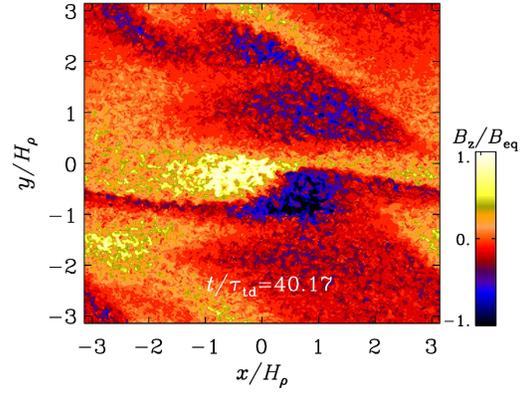}
\end{center}\caption{
$B_z/\Beq$ in the $xy$ plane through $z/H_\rho=\pi$ for
DNS Run~Rn6 with non-helically forced rotating turbulence.
}\label{pbz_xy56}\end{figure}

We have performed a number of runs with varying Coriolis number, $\Co$.
We also vary colatitude $\theta$ (Runs~Rn2--4).
Our simulations show that for fast rotation ($\Co>5.6$)
and for $\theta=0$ (Runs~Rn6 and Rn7),
the self-generated kinetic helicity leads to an $\alpha^2$ dynamo.
At a later stage, bipolar magnetic structures form.
\FFig{pxyav56} presents the time evolution of kinetic and magnetic helicities
together with the horizontal components of the magnetic field.
One can see the production of negative kinetic helicity in almost all
of the entire domain (see the upper left panel of \Fig{pxyav56}).
Magnetic helicity, however, has both negative and positive signs and has
a non-zero value in the lower part of the domain (see the upper right panel
of \Fig{pxyav56}).
In the lower row of \Fig{pxyav56}, we present the butterfly diagram of
$B_x/\Beq(z)$ (left), and $B_y/\Beq(z)$ (right).
There is a clear phase shift between these two field components,
similar to what one expects from a Beltrami-type magnetic
field driven by $\alpha^2$ dynamo.
This large-scale magnetic field reaches the surface
and bipolar magnetic spots are formed; see \Fig{pbz_xy56}.

\section{Conclusions}
\label{Conclusions}

In this work we have compared DNS of
helically forced turbulence in a strongly
stratified layer with corresponding MFS.
Compared with earlier DNS \citep{MBKR14,Jabbari16}, we have considered here
a one-layer model and have shown that this simpler case also
leads to the formation of sharp bipolar structures at the surface.
Larger values of $\Rm$ result in more complex spatio-temporal behavior,
while rotation (with $\Co\la1$)
and the scale separation ratio have
only minor effects.
Both aspects confirm similar findings for our earlier two-layer model.

The results of our MFS are generally in good qualitative agreement with
the DNS.
The MFS without parameterized NEMPI
(i.e., neglecting the turbulence effects on the Lorentz force)
demonstrate that the formation
of sharp structures at the surface occurs
predominantly due to the nonlinear effects associated with the
mean Lorentz force of the dynamo-generated magnetic field,
provided the dynamo number is at least 2.5 times supercritical.
This results in converging flow structures and downdrafts in equivalent
locations both in DNS and MFS.
For smaller dynamo numbers, when the field strength is below equipartition,
NEMPI can operate and form bipolar regions, as was shown in earlier
DNS \citep{WLBKR13,WLBKR16}.
Comparing MFS without and with inclusion of the parameterization of NEMPI
by replacing the mean Lorentz force with the effective Lorentz force
in the Navier-Stokes equation, we found that the formation of bipolar
magnetic structures in the case of NEMPI is also accompanied by
downdrafts, especially in the upper parts of the computational domain.

In this connection, we recall that our system lacks the effects of thermal
buoyancy, so our downdrafts are distinct from those in convection.
In the Sun, both effects may contribute to driving convection,
especially on the scale of supergranulation.
However, from convection simulations with and without magnetic fields,
no special features of magnetically driven downflows have been seen
\citep{KBKKR16}.

Finally, we have considered nonhelically forced turbulence, but now with
sufficiently rapid rotation which, together with density stratification,
leads to an $\alpha$ effect that is supercritical for the onset of
dynamo action.
Even in that case we find the formation of sharp bipolar structures.
They begin to resemble the structures of bipolar regions in the Sun.
Thus, we may conclude that the appearance of bipolar structures at the
solar surface may well be a generic feature of a large scale dynamo some
distance beneath the surface of a strongly stratified domain.

As a next step, it will be important to consider more realistic modeling
of the large-scale dynamo.
This can be done in global spherical domains with differential rotation,
which should lead to preferential east-west alignment of the bipolar
structures.
In addition, the effects of convectively-driven turbulence would be
important to include.
This would automatically account for the possibility of thermally driven
downflows, in addition to just magnetically driven flows.
In principle, this has already been done in the many global dynamo
simulations performed in recent years \citep{BMBBT11,KMB12,FF14,Hotta},
but in most of them the stratification was not yet strong
enough and the resolution insufficient to resolve small
enough magnetic structures at the surface.

The spontaneous formation of magnetic surface structures from a
large-scale $\alpha^2$ dynamo by strongly stratified thermal convection
in Cartesian geometry has recently also been studied by \cite{MS16}.
They found that large-scale magnetic structures are formed at the
surface only in cases with strong stratification.
However, in many other convection simulations, the scale separation
between the integral scale of the turbulence and the size of the domain
is not large enough for the formation of sharp magnetic structures.
One may therefore hope that future simulations will not only be more
realistic, but will also display surface phenomena that are closer
to those observed in the Sun.

\section*{Acknowledgements}
We thank Dhrubaditya Mitra for useful discussions regarding this work.
It was supported in part by
the Swedish Research Council Grants No.\ 621-2011-5076 (AB, SJ),
2012-5797 (AB), Australian Research Council's Discovery Projects 
funding scheme project No. \ DP160100746 (SJ), and the
Research Council of Norway under the FRINATEK grant 231444 (AB, IR).
It was also supported by The Royal Swedish Academy of Sciences grant
No.\ AST2016-0026 (SJ).
We acknowledge the allocation of computing resources provided by the
Swedish National Allocations Committee at the Center for Parallel
Computers at the Royal Institute of Technology in Stockholm.
This work utilized the Janus supercomputer, which is supported by the
National Science Foundation (award number CNS-0821794), the University
of Colorado Boulder, the University of Colorado Denver, and the National
Center for Atmospheric Research. The Janus supercomputer is operated by
the University of Colorado Boulder.


\label{lastpage}
\end{document}